\documentclass[aps,amsfonts,floatfix]{revtex4}	

\usepackage{latexsym,amssymb}
\usepackage[dvips]{graphics}
\usepackage[dvips]{color}
\usepackage{graphics,psfrag,graphicx,subfigure,bbm}
\usepackage{amsmath,amsthm,amscd} 

\newcommand{\Rset}{{\mathbb R}}
\newcommand{\Tset}{{\mathbb T}}
\newcommand{\ie}{{i.e.\ }}
\newcommand{\eg}{{e.g.\ }}

\newcommand{\beq}{\begin{equation}}
\newcommand{\eeq}{\end{equation}}
\newcommand{\bea}{\begin{eqnarray}}
\newcommand{\eea}{\end{eqnarray}}
\newcommand{\ba}{\begin{array}}
\newcommand{\ea}{\end{array}}
\newcommand{\bean}{\begin{eqnarray*}}
\newcommand{\eean}{\end{eqnarray*}}
\newcommand{\isubfig}[1]{\subfigure{#1}\nonumber}
\newcommand{\xyvect}[2]{\left(\begin{matrix} #1 \\ #2 \end{matrix}\right)}

\begin{document}
\title{Renormalization and destruction of $1/\gamma^2$ tori in the
  standard nontwist map}
\author{ A.~Apte, A.~Wurm and P.J.~Morrison}
\affiliation{Department of Physics and Institute for Fusion Studies\\
University of Texas at Austin, Austin, Texas 78712}
\date{\today}
\begin{abstract}
Extending the work of del-Castillo-Negrete, Greene, and Morrison,
Physica D {\bf 91}, 1 (1996) and {\bf 100}, 311 (1997) on the
standard nontwist map, the breakup of an invariant torus
with winding number equal to the inverse golden mean squared is
studied. Improved numerical techniques provide the greater accuracy
that is needed for this case. The new results are interpreted within
the renormalization group framework by constructing a renormalization
operator on the space of commuting map pairs, and by studying the
fixed points of the so constructed operator.
\end{abstract}
\maketitle

{\bf 
In recent years, area-preserving maps that violate the twist condition
locally in phase space have been the object of interest in several
studies in physics and mathematics. These \emph{nontwist} maps show
up in a variety of physical models. An important problem from the
physics point of view is the understanding of the breakup of invariant
tori, which show remarkable resilience in the region where the
twist condition is violated, called \emph{shearless tori}. In terms of
the physical system modelled, these tori represent transport barriers, and their
breakup corresponds to the transition to global chaos. Mathematically,
nontwist maps present a challenge since the standard proofs of
celebrated theorems in the theory of area-preserving maps rely heavily
on the twist condition. In this paper, we study the breakup of the
shearless torus with winding number \boldmath{$1/\gamma^2$}, where
\boldmath{$\gamma$} is the golden mean. This torus serves as a test
case for improved techniques we developed. At the point of breakup the
shearless torus exhibits universal scaling behavior which leads to a
renormalization group interpretation.} 

\section{Introduction}\label{sec:intro}
In this paper we consider the {\it standard nontwist map} (SNM) $M$, as
introduced in Ref.~\onlinecite{diego2}:
\bea
x_{n+1} & = & x_n + a \left(1-y^2_{n+1}\right)\nonumber\\[0.1in]
y_{n+1} & = & y_n - b \sin\left(2\pi x_n\right),
\label{eq:stntmap}
\eea
where $(x,y)\in\Tset\times\Rset$, $a\in(0,1)$, and
$b\in(-\infty,\infty)$. The map $M$ is {\it area-preserving} and
violates the {\it twist condition}
\beq
\frac{\partial x_{i+1}\left(x_i,y_i\right)}{\partial y_i}
 \neq 0\qquad\qquad \forall (x_i,y_i),
\eeq
along a curve in phase space, which has been recently called the
{\it nonmonotone curve}.\cite{petrisor1} Traditionally, most studies of
area-preserving maps have dealt with the {\it twist} case, but in
recent years more and more research has been focused on the nontwist
case. 

Applications of nontwist maps occur in many fields, for example, the
study of magnetic field lines in toroidal plasma devices (see \eg
Refs.~\onlinecite{horton1,balescu}), in celestial mechanics,\cite{kyner}
fluid dynamics\cite{diego2} and atomic physics.\cite{chandre2}
It has been shown\cite{dullin,vander} that nontwist regions appear
generically in area-preserving maps that have a tripling bifurcation
of an elliptic fixed point.
In addition to these applications, the map is quite interesting from
a mathematical standpoint because many important theorems in the theory
of area-preserving maps assume the validity of the twist condition,
\eg the KAM theorem and the Poincare-Birkhoff theorem. The SNM can
serve as a model for the development of new proofs. Up to now, only a
few mathematical results exist for nontwist maps (see 
\eg Ref.~\onlinecite{delshams,franks,petrisor1,simo}).

We continue the work of del-Castillo-Negrete, Greene and
Morrison,\cite{diego4,diego5} who studied the breakup of the
shearless invariant torus with winding number $1/\gamma$, where
$\gamma=(1+\sqrt{5})/2$ is the golden mean. We present the analysis
of the breakup of the shearless invariant torus with winding number
(in continued fraction representation)
\beq
\omega=[0,2,1,1,\ldots]=1/\gamma^2.
\eeq
Because this winding number is a noble number (its continued fraction expansion
ends with $[1,1,1,\ldots]$), the behavior of the residues of the
approximating periodic orbits is expected to be the same as in the
$1/\gamma$ case, \ie we
should find the same fixed point of the renormalization group
operator with the same unstable eigenvalues that were found in 
Ref.~\onlinecite{diego5}. But, the form of the renormalization group operator,
which is defined later in Sec.~\ref{sec:reng}, is
different from the $1/\gamma$ case.
Also, the region of parameter space we study is different.
Additionally, since the periods of approximating
periodic orbits are bigger than those for the $1/\gamma$ case, the
present work serves as a test case for improved numerical techniques
described later in Sec.~\ref{ssec:num}.

A different approach, which yields rough parameter
values for the breakup of invariant tori, was used by Shinohara
and Aizawa in Ref.~\onlinecite{shin1}, who showed that a shearless invariant
torus crosses the $x$-axis at two points,\cite{note1} 
$x_A = a/2-1/4$ and $x_B = a/2+1/4$.
\begin{figure}[!t]
\centering
\includegraphics[angle=270,width=0.48\textwidth]{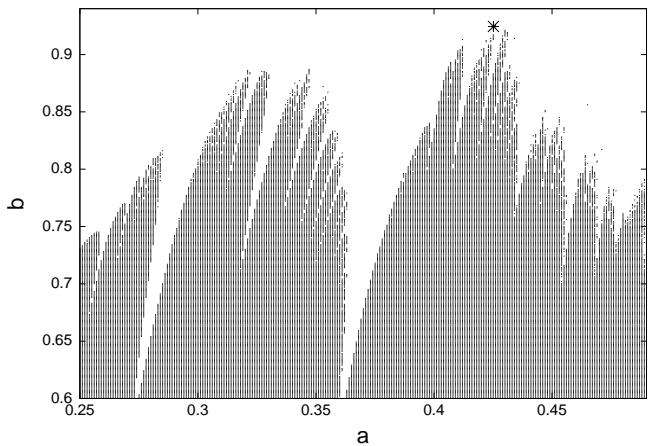}
\caption{Parameter space around the critical point (marked by $\ast$)
of the $\omega=1/\gamma^2$-shearless curve, showing
the points for which shearless invariant tori exist.}
\label{fig:stntshin2}
\end{figure}
For a given $(a,b)$, a point on the shearless torus,
$(a/2+1/4,0)$, is iterated many times (we used $10^6$). If the $y$ value
stays below a threshold (we used $|y|<0.52$), it is assumed that the
shearless curve exists and the point is plotted. Figure
\ref{fig:stntshin2} depicts our duplication of their procedure. 

We see that the critical point for the
$1/\gamma^2$ shearless curve (indicated by $\ast$) lies on the
boundary in Fig.~\ref{fig:stntshin2}. Thus,
the boundary points of Fig.~\ref{fig:stntshin2} represent the {\it
critical function} for the SNM. This is a generalization of 
the definition of the critical function for the standard twist map
(see \eg Ref.~\onlinecite{stark}), which has only one parameter \eg $k$. The
critical function in the twist case is then defined as
$k_c(\omega)$. Here, we have two parameters, but the shearless
invariant torus of a given winding number $\omega$ exists only for
parameter values belonging to a curve
$\left(a,b\left(a;\omega\right)\right)$ in the parameter space. Thus,
we can define the critical function by the critical points on each of
those curves by $\left(a_c,b\left(a_c;\omega\right)\right)$. By
finding the critical points for many other winding numbers (both
nobles and non-nobles), we hope to find a more accurate critical
function curve than the one shown in Fig.~\ref{fig:stntshin2}.

In Sec.~\ref{sec:review}, we review some basic properties of the
SNM. The detailed breakup of the shearless invariant torus with
winding number $1/\gamma^2$ is presented in Sec.~\ref{sec:neww},
which also contains a discussion of the numerical procedures
involved. In Sec.~\ref{sec:reng}, we interpret the results within
the framework of the renormalization group. Section \ref{sec:concl}
a summary and some directions of future research.

\section{Review of residue criterion and standard nontwist map}
\label{sec:review}
In this section, we give a brief review of some basic concepts of the theory of
area-preserving maps in the context of the SNM. For a more in-depth
discussion the reader is referred to Ref.~\onlinecite{diego4} and
references therein. 

\subsection{Periodic orbits and residue criterion}\label{ssec:po}
Since the pioneering work of Greene,\cite{greene2} periodic orbits
have proven to be very useful for studying the breakup of invariant
tori in area-preserving maps. Below are some standard definitions.

An {\it orbit} of an area-preserving map $M$ is a sequence of points 
$\left\{\left(x_i,y_i\right)\right\}$ such that $M\left(x_i,y_i\right)
= \left(x_{i+1},y_{i+1}\right)$. 
The {\it winding number} $\omega$ of an orbit is defined as
$\omega = \lim_{i\to\infty} (x_i/i)$
when the limit exists. Here the $x$-coordinate is ``lifted'' from
$\Tset$ to $\Rset$.
A {\it periodic orbit} of period $n$ is a sequence of $n$ points
$\left\{\left(x_i,y_i\right)\right\}_{i=1}^n$, such that 
$M^n \left( x_i, y_i\right) = \left( x_i+m, y_i\right)$
for all $i=1,\ldots,n$,
and $m$ is an integer. Periodic orbits have rational winding numbers
$\omega=m/n$.
An {\it invariant torus} is an orbit with irrational winding
number that covers densely a one-dimensional set in phase space. Of
particular importance are the invariant tori that wind around the
$x$-domain because, in two-dimensional maps, they act as transport
barriers.

The linear stability of a periodic orbit is determined by the
value of its {\it residue},\cite{greene2} $R$, which is defined as
$R:=\left[2-\mbox{Tr}(L)\right]/4$. Here, $L$ is the map $M^n$
linearized about the periodic orbit of interest and Tr denotes the
trace. If $0<R<1$, the orbit is stable or elliptic; if $R<0$ or $R>1$,
it is unstable or hyperbolic; in the degenerate cases $R=0$ and $R=1$,
it is parabolic.

Periodic orbits can be used to systematically approximate invariant
tori.\cite{greene2} The method is based on the observation that given
a sequence of rational numbers $\left\{m_i/n_i\right\}$ whose limit is
$\omega$, the sequence of periodic orbits with winding numbers
$\left\{m_i/n_i\right\}$ approaches the invariant torus with winding
number $\omega$ in phase space. It is important to find the ``best''
possible sequence, \ie the sequence that converges to $\omega$ the
fastest. The elements of the best possible sequence (see \eg
Ref.~\onlinecite{khinchin}) are the
convergents that are obtained from successive truncations of the
continued fraction expansion of $\omega$.

The {\it residue criterion\/}\cite{greene2} can be stated as follows:
Consider an invariant torus with winding number $\omega$. Let $\{m_i/n_i\}$ 
be the sequence of convergents approximating $\omega$, and $R_i$ the
residues of their corresponding periodic orbits.
\begin{enumerate}
\item If $\lim_{i\to\infty} |R_i| = 0$, the invariant torus exists.
\item If $\lim_{i\to\infty} |R_i| =\infty$, the invariant torus is
destroyed.
\item At the boundary in parameter space between those two limits, the
invariant torus is at the threshold of destruction and the residues either
converge to a constant, non-zero value, or there are convergent subsequences.
\end{enumerate}

This criterion is based on the idea that the destruction of an invariant
torus is caused by the de-stabilization of nearby periodic orbits.
The residue criterion has been used successfully in many cases to
predict with high precision the threshold for the destruction of 
invariant tori. Several theorems have been proved
 that lend mathematical support to the criterion.\cite{falc,mackay2}

The numerical search for periodic orbits is difficult because, in principle,
it is a two-dimensional root finding problem. However, the task is
considerably simplified for {\it reversible
  maps},\cite{greene2,devogel} which are maps that can be factored
as $M = I_1\circ I_0$,
where $I_{0,1}$ are {\it involution} maps that satisfy $I_1^2 =I_0^2 =1$.
The sets of fixed points of the involution maps,
$\Gamma_{0,1} = \left\{ (x,y)| I_{0,1} (x,y) = (x,y)\right\}$,
are one-dimensional sets, called {\it symmetry lines} of the map. Once
we know $\Gamma_{0,1}$,  the search for periodic orbits is reduced to
a one-dimensional root finding problem, as explained below in
Sec.~\ref{ssec:spo}.

\subsection{Standard nontwist map}\label{ssec:std_nt}

The SNM is reversible. The symmetry lines $\Gamma_0$, composed of
fixed points of $I_0$ are $s_1 = \left\{(x,y) |x=0\right\}$ and
$s_2 = \left\{(x,y) |x=1/2\right\}$.
The symmetry lines $\Gamma_1$, composed of fixed points of $I_1$ are
$s_3 = \left\{(x,y) |x=a\left(1-y^2\right)/2\right\}$ and
$s_4 = \left\{(x,y) |x=a\left(1-y^2\right)/2+1/2\right\}$.

A major difference between the standard nontwist map and twist
maps is that there are two periodic orbits, if they exist, with the
same winding number on each symmetry line. This can be seen easily
in the integrable case. For $b=0$, the $m/n$ periodic orbits on the
$s_1$ symmetry line are located at
\beq
(x,y) = \left( 0,\pm\sqrt{1-(m/n)/a}\right).
\label{eq:bzero}
\eeq
We will call the orbit with the bigger (smaller) $y$-coordinate the
{\it up} ({\it down}) periodic orbit.

The SNM is also invariant with respect to the transformation
\beq
T\left(x,y\right) = \left(x+1/2,\,-y\right).
\label{eq:tdef}
\eeq
The coordinates of the up and down periodic orbits on the symmetry
lines $s_i$, denoted by $(x_{ui},y_{ui})$ and $(x_{di},y_{di})$
respectively, are related by this symmetry as follows:
\bea
(x_{d2},y_{d2}) = T\left((x_{u1},y_{u1})\right), \qquad \qquad 
                  (x_{u2},y_{u2})  =  T\left((x_{d1},y_{d1})\right), \nonumber\\
(x_{d4},y_{d4}) = T\left((x_{u3},y_{u3})\right), \qquad \qquad
                  (x_{u4},y_{u4})  =  T\left((x_{d3},y_{d3})\right).
\label{eq:tmap}
\eea
Therefore, it is actually enough to compute periodic orbits on
$s_1$ and $s_3$, since the orbits along the other symmetry lines
can be obtained from (\ref{eq:tmap}).

\subsection{Periodic orbit collisions and bifurcation curves}
\label{ssec:snmpocoll}

Periodic orbits in the SNM can undergo a particular
form of bifurcation that
occurs when the up and down periodic orbits of the same winding number
meet on the symmetry line. These collisions were detected numerically
in Refs.~\onlinecite{stix,howard} and \onlinecite{diego4}.
Further studies of this bifurcation can be found in
Refs.~\onlinecite{petrisor1,simo}. 

\begin{figure}[ht]
\centering
\includegraphics[angle=270,width=0.48\textwidth]{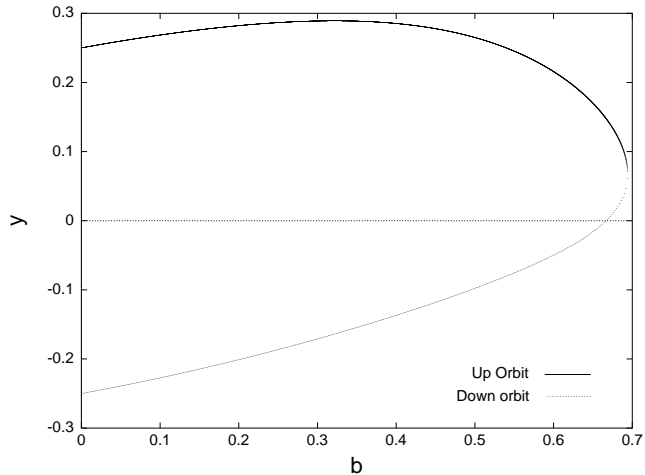}
\caption{Plot of behavior of the up and down periodic
orbit of winding number $3/8$ for increasing $b$-values at $a=0.4$. The
vertical axis shows the $y$-coordinates of the orbits along $s_1$.}
\label{fig:ybcurve}
\end{figure}

From (\ref{eq:bzero}) it follows that, for a given $a$, only periodic
orbits with $m/n < a$ exist at $b=0$. As the value of $b$ increases,
the up and down orbits approach each other and at the bifurcation
value, they collide and annihilate each other. For higher values of $b$, both
orbits no longer exist. Figure \ref{fig:ybcurve} illustrates the
behavior of periodic orbits as we increase $b$ from $b=0$. Here the
$y$-coordinates of the $m/n = 3/8$ periodic orbits on $s_1$ is shown as
a function of $b$ for the fixed value of $a=0.4$.
\begin{figure}[ht]
\center
\includegraphics[angle=270,width=.48\textwidth]{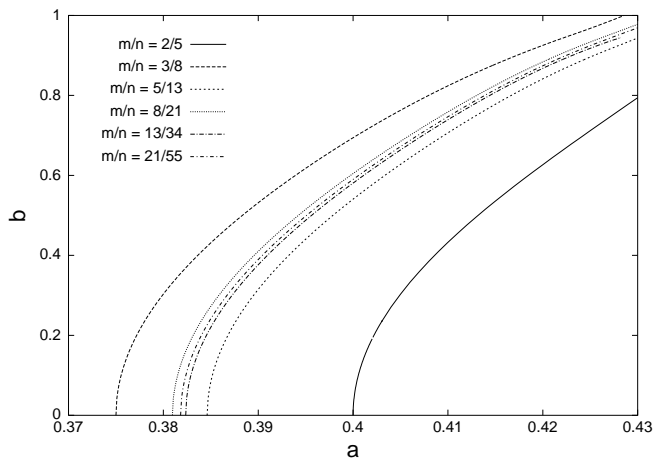}
\caption{\label{fig:bifurcurves} Bifurcation curves for several
convergents of $1/\gamma^2$.}
\end{figure}

Based on these numerical observations, the notion of a bifurcation
curve in parameter space was defined in Ref.~\onlinecite{diego4}.
The {\it $m/n$-bifurcation curve
$b=\Phi_{m/n}(a)$} is the set of $(a,b)$ values for which the $m/n$ up
and down periodic orbits are at the point of collision.
The main property of this curve is that for $(a,b)$ values below 
$b=\Phi_{m/n}(a)$, the $r/s$ periodic orbits, with $r/s < m/n$ exist.
Thus, $m/n$ is the maximum winding number for parameter values along
the $m/n$-bifurcation curve.

The idea of approximating invariant tori with irrational winding
numbers by periodic orbits is used to define the bifurcation curve for an
invariant torus as follows:\cite{diego4}
The {\it $\omega$-bifurcation curve $b=\Phi_\omega(a)$} for an
irrational $\omega$ is the set of $(a,b)$ values such that
$b=\Phi_\omega(a)=\lim_{i\to\infty} \Phi_{m_i/n_i}(a)$, where
$\Phi_{m_i/n_i}(a)$ is the $m_i/n_i$-bifurcation curve and
$\{m_i/n_i\}$ are the convergents of $\omega$.
For $(a,b)$ points along the $\omega$-bifurcation curve the invariant
torus with irrational winding number $\omega$ is the curve of maximum
winding number and is called {\it shearless}.
Figure \ref{fig:bifurcurves} depicts the bifurcation curves for
several convergents of $1/\gamma^2$. This figure also makes it
plausible that the limit in the above definition exists.

\section{Breakup of Torus with  $\omega=1/\gamma^2$}\label{sec:neww}

In this section, we present the analysis of the breakup of the
shearless invariant torus in the standard nontwist map with winding
number $\omega=1/\gamma^2$. Tables \ref{tab:conv} and \ref{tab:conv2}
list the convergents used for these calculations. For more details
see Ref.~\onlinecite{wurm}.

\begin{table}[ht]
\begin{center}
\begin{tabular}{|rl||rl|}\hline
&&&\\[-2mm]
$[i]$ & $F_i/F_{i+2}$ & $[i]$ & $F_i/F_{i+2}$\\[2mm]
\hline
$[1]$  & 1/3       & $[19]$ & 6765/17711\\
$[3]$  & 3/8       & $[21]$ & 17711/46368\\
$[5]$  & 8/21      & $[23]$ & 46368/121393\\
$[7]$  & 21/55     & $[25]$ & 121393/317811\\
$[9]$  & 55/144    & $[27]$ & 317811/832040\\
$[11]$ & 144/377   & $[29]$ & 832040/2178309\\
$[13]$ & 377/987   & $[31]$ & 2178309/5702887\\
$[15]$ & 987/2584  & $[33]$ & 5702887/14930352\\
$[17]$ & 2584/6765 & $[35]$ & 14930352/39088169 \\
\hline
\end{tabular}
\caption{\label{tab:conv}Some of the convergents of
  $\omega=[0,2,1,1,\ldots]$ for which the periodic orbits still
  exist at criticality.}
\end{center} 
\end{table}

\begin{table}[ht]
\begin{center}
\begin{tabular}{|rl||rl|}\hline
&&&\\[-2mm]
$[i]$ & $F_i/F_{i+2}$ & $[i]$ & $F_i/F_{i+2}$\\[2mm]
\hline
$[2]$  & 2/5       & $[18]$ & 4181/10946\\
$[4]$  & 5/13      & $[20]$ & 10946/28657\\
$[6]$  & 13/34     & $[22]$ & 28657/75025\\
$[8]$  & 34/89     & $[24]$ & 75025/196418\\
$[10]$ & 89/233    & $[26]$ & 196418/514229\\
$[12]$ & 233/610   & $[28]$ & 514229/1346269\\
$[14]$ & 610/1597  & $[30]$ & 1346269/3524578\\
$[16]$ & 1597/4181 & $[32]$ & 3524578/9227465\\
\hline
\end{tabular}
\caption{\label{tab:conv2}Some of the convergents of
  $\omega=[0,2,1,1,\ldots]$ for which the periodic orbits do not exist
  at criticality.}
\end{center} 
\end{table}

\subsection{Numerical methods}\label{ssec:num}

The computational steps necessary to find the critical point
and the residue behavior of the approximating periodic
orbits are as follows:
\begin{enumerate}
\item Find a good approximation to the $1/\gamma^2$-bifurcation
curve in $(a,b)$-space using the bifurcation curves for its
convergents.
\item Along this bifurcation curve, find the up and down
periodic orbits on the symmetry line $s_1$ that approximate
the invariant torus, and compute their residues.
\item Locate the $(a,b)$ point along the curve at which 
the residues exhibit critical behavior.
\item Find the residues of the periodic orbits at 
criticality along the remaining symmetry lines.
\item Find the eigenvalues of the unstable eigenmodes
of the renormalization group operator. The
details of how to do this depend crucially on the type of critical
scaling behavior that is exhibited by the residues.
\end{enumerate}

\subsubsection{Searching for periodic orbits}\label{ssec:spo}

Periodic orbits on the symmetry lines can be be computed relatively
easily for reversible maps using the following result:\cite{diego4}
If $(x,y) \in \Gamma_{0,1}$ then $M^n(x,y)=(x,y)$ if and only if
$M^{n/2}(x,y) \in \Gamma_{0,1}$ (for $n$ even) or $M^{(n\pm 1)/2}(x,y) \in
\Gamma_{1,0}$ (for $n$ odd).
Thus, for example, periodic orbits with odd period $n$
on the $s_1$ symmetry line can be obtained by looking for points 
on $s_1$ that are mapped to $s_3$ or $s_4$ after
$(n+1)/2$ iterations. This can be implemented as a one-dimensional root
finding problem by considering the zeros of the function
$F(y) = \sin \left[ 2\pi\left(\hat{x}-a\left(1-\hat{y}^2\right)/2\right)
\right]$, where
$\left( \hat{x},\hat{y}\right):= M^{(n+1)/2}(0,y)$.
The sine function is used to eliminate the difference between $s_3$ and
$s_4$. Similar ideas can be applied to find other orbits.

\subsubsection{Finding $m/n$-bifurcation curves}\label{sssec:fbif}

Recall that the bifurcation curve $\Phi_{m/n}(a)$ of a periodic orbit of
winding number $m/n$ was defined in Sec.~\ref{ssec:snmpocoll} to be
the set of points $(a,b)$, at which the up and down periodic orbits of
winding number $m/n$ collide along the $s_1$ symmetry line. Thus, at a
given value of $a$, the function $F(y)$ has two
roots for $b<\Phi_{m/n}(a)$, no roots (locally) for $b>\Phi_{m/n}(a)$
and a single root, which is also an extremum, for $b=\Phi_{m/n}(a)$.
We thus search for the zero of the extremum of $F(y)$ as $b$ is varied.

To find the whole (or large portions) of a bifurcation curve, we
use the monotonic nature of the curve (see Fig.~\ref{fig:bifurcurves})
as follows: Given a point $(a_1,b_1)$ on the
bifurcation curve \ie $b_1=\Phi_{m/n}(a_1)$, we increase $a$ by a fixed
amount to $a_2=a_1+a_{\mbox{\scriptsize step}}$. We then start at the point
$(a_2,b_1)$ and increase $b$ until we reach $b_2=\Phi_{m/n}(a_2)$.
To make sure that we are finding the correct bifurcation curve, we 
start searching $(a,b)$-space at $(a,b)=(m/n,0)$. Even then,
the steps in $a$ cannot be taken to be too large. Experience has shown that
steps in $a$ of $1\times 10^{-5}$ or $1\times 10^{-6}$ are safe.
This method is unfortunately very slow because the part of the curve
at small $b$ values is very steep and the interesting (near critical)
part of the curve is far away from the $b=0$ limit.

We managed to drastically improve the speed of these calculations by
using the following ideas:
\begin{enumerate}
\item Numerical evidence strongly suggests that a bifurcation curve
is smooth and monotonically increasing, although it is not proved
mathematically.\cite{note2} So we
use linear extrapolation from two previous points to find the new
value of $b$ around which to search for the bifurcation point. It was
found that any higher order extrapolation did not improve the
algorithm further.
\item To find bifurcation curves for periodic orbits
with very large periods (\eg of the order of several
million) the following procedure is used: Starting at the bifurcation
curve of a smaller period, we increase $b$ until the bifurcation curve
of the higher period is reached. The advantage of this procedure is
that we do not need to do the extremely time consuming calculations of
the bifurcation curves for very high period orbits starting at $b=0$,
but rather we can search for them near the region of interest.
\end{enumerate}

\subsubsection{Finding $1/\gamma^2$-bifurcation curve}\label{sssec:wbif}

Recall that the $1/\gamma^2$-bifurcation curve was defined as the limit of
$m_i/n_i$ bifurcation curves, where $m_i/n_i$ are convergents of
$1/\gamma^2$. It was numerically observed that close to criticality,
this limit is approached in accordance with the following
scaling relation:\cite{diego4} 
\beq
\Phi_{[n+1]}(a) =  \Phi_{1/\gamma^2}(a) + B_n(a)\; \nu_1^{n/12},
\label{eq:scalrel}
\eeq
where the $\Phi_{[n]}(a)$ denotes the bifurcation curve of the
periodic orbit with winding number $[n]=F_n/F_{n+2}$, $\nu_1$ is a
number to be determined later, and $B_n(a)$ is a period-twelve function,
\ie $B_{n+12}(a)=B_n(a)$.

If Eq.~(\ref{eq:scalrel}) holds, it follows that for fixed $a$
\beq
\Phi_{1/\gamma^2} = \lim_{n\to\infty} \frac{\Phi_{[n+1]} \Phi_{[n+12]}-
\Phi_{[n]} \Phi_{[n+13]}}{\left(\Phi_{[n+1]}-\Phi_{[n]}\right)-
\left(\Phi_{[n+13]}- \Phi_{[n+12]}\right)}.
\label{eq:scalrel2}
\eeq
We obtained the $1/\gamma^2$-bifurcation curve using $n=19$ in
Eq.~(\ref{eq:scalrel2}), \ie using the bifurcation curves for $[32]$,
$[31]$, $[20]$ and $[19]$ (see Tables \ref{tab:conv} and
\ref{tab:conv2}).

Now one can justify {\it a posteriori} the use of Eq.~(\ref{eq:scalrel}).
Solving Eq.~(\ref{eq:scalrel}) with $a=a_c$ for $\nu_1$ yields:
\beq
\nu_1 = \lim_{n\to\infty}
        \left(\frac{\Phi_{[n+13]}\left(a_c\right)-b_c}
                   {\Phi_{[n+1]}\left(a_c\right)-b_c}\right),
\eeq
and
\beq
B_n(a_c) = \left( \Phi_{[n+1]}\left(a_c\right)-b_c\right)\; \nu_1^{-n/12},
\eeq
where $(a_c,b_c)$ is the critical point for breakup of the shearless
$1/\gamma^2$ invariant torus \ie $b_c=\Phi_{1/\gamma^2}(a_c)$.
We found that $\nu_1^{-1/12}=2.678$. Some numerical 
evidence for the periodicity of $B_n(a_c)$ is given in
Table \ref{tab:bn}.

\begin{table}[ht]
\begin{center}
\begin{tabular}{|rr|rr|}\hline
&&&\\[-2mm]
$n$ &$B_n(a_c)$ & $n$  & $B_n(a_c)$\\[2mm]
\hline
$15$  & -0.4865 & $27$ & -0.4865\\
$17$  & -0.7090 & $29$ & -0.7078\\
$18$  &  0.5019 & $20$ &  0.5028\\
$19$  & -0.3901 & $31$ & -0.3887\\
\hline
\end{tabular}
\caption{\label{tab:bn} Period-twelve behavior
of the scaling function $B_n(a_c)$.} 
\end{center}
\end{table}

\subsection{Results}\label{ssec:results}

In this subsection, we present the results of our computations.

\subsubsection{Residue behavior at criticality}

We computed bifurcation curves up to $[32]=3524578/9227465$ and found the
critical points along them, \ie the parameter values along those curves
for which the residues of approximating periodic orbits neither
converge to zero nor diverge to infinity.
\begin{figure}[!ht]
\centering
    \isubfig{
	 \includegraphics[angle=270,scale=2,width=.8\textwidth]{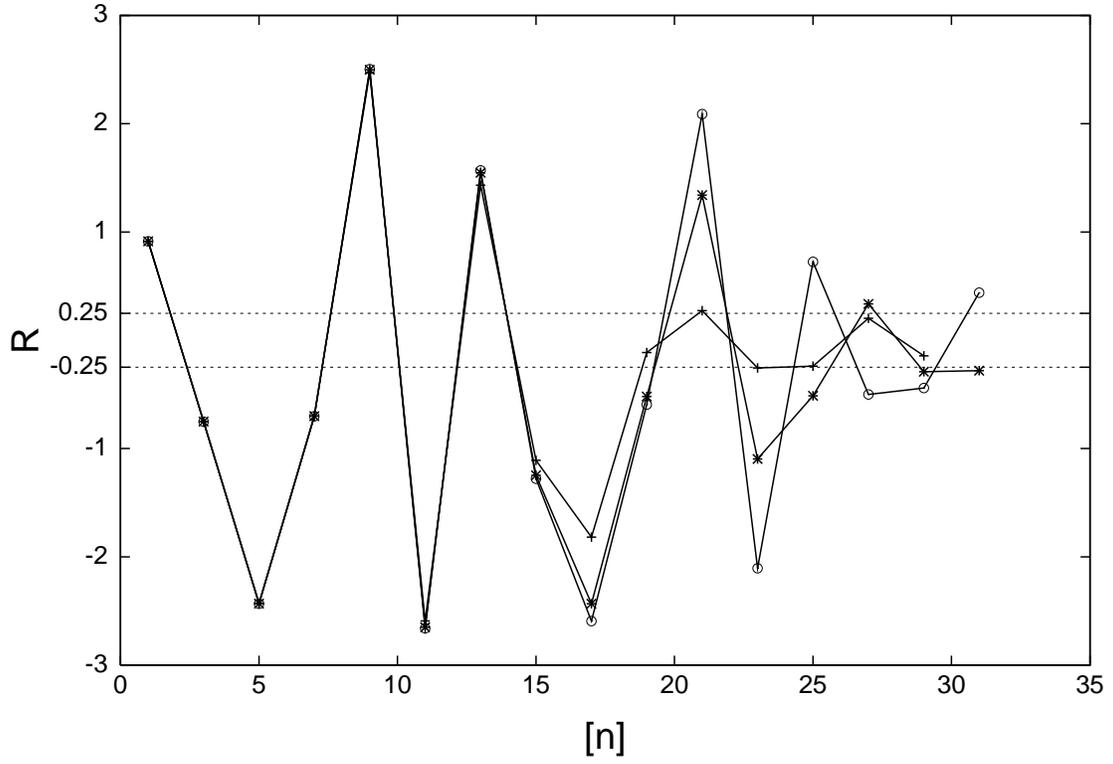}
    }
    \isubfig{			       
	 \includegraphics[angle=270,scale=2,width=.8\textwidth]{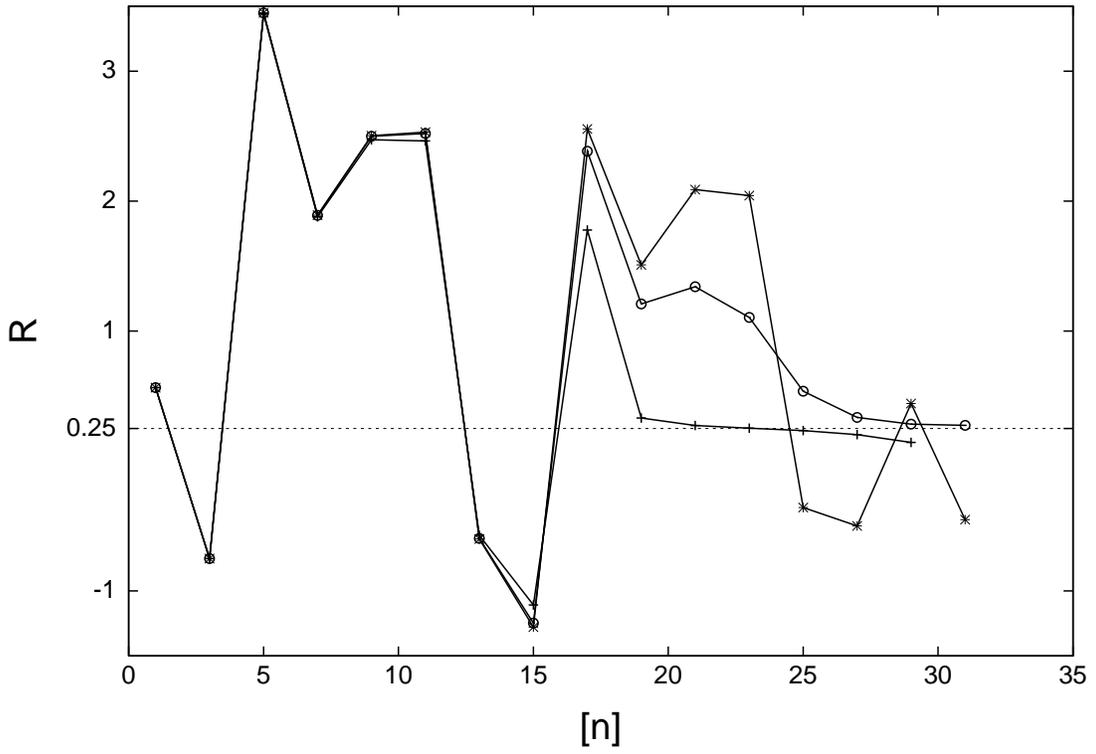}
    }
\caption{Residue behavior of the up (top figure) and down (bottom
figure) periodic orbits on $s_1$ at the critical points on bifurcation
curves of $[20]$ ($+$), $[24]$ ($\circ$)and $[28]$ ($\ast$).}
\label{fig:updownorb2}
\end{figure}
Figure \ref{fig:updownorb2} shows the critical residue behavior of the
up and down periodic orbits on the symmetry line $s_1$ along several
different bifurcation curves. For lower period bifurcation curves, the
residues first show signs of a six-cycle (to be discussed later in
greater details), but then converge to $\left|R_i\right|\approx
0.25$. This is because the invariant torus we are studying is not quite
shearless. Thus we see the same behavior of the
residues as in the case of a twist map. As we proceed to higher period
bifurcation curves, the behavior of the residues of the approximating
periodic orbits found along the $s_1$ symmetry line resembles more and
more a six-cycle. A renormalization group interpretation of these
results is given in Sec.~\ref{sec:reng}.

\begin{figure}[ht]
\centering
    \isubfig{
	 \includegraphics[angle=270,scale=2,width=.45\textwidth]{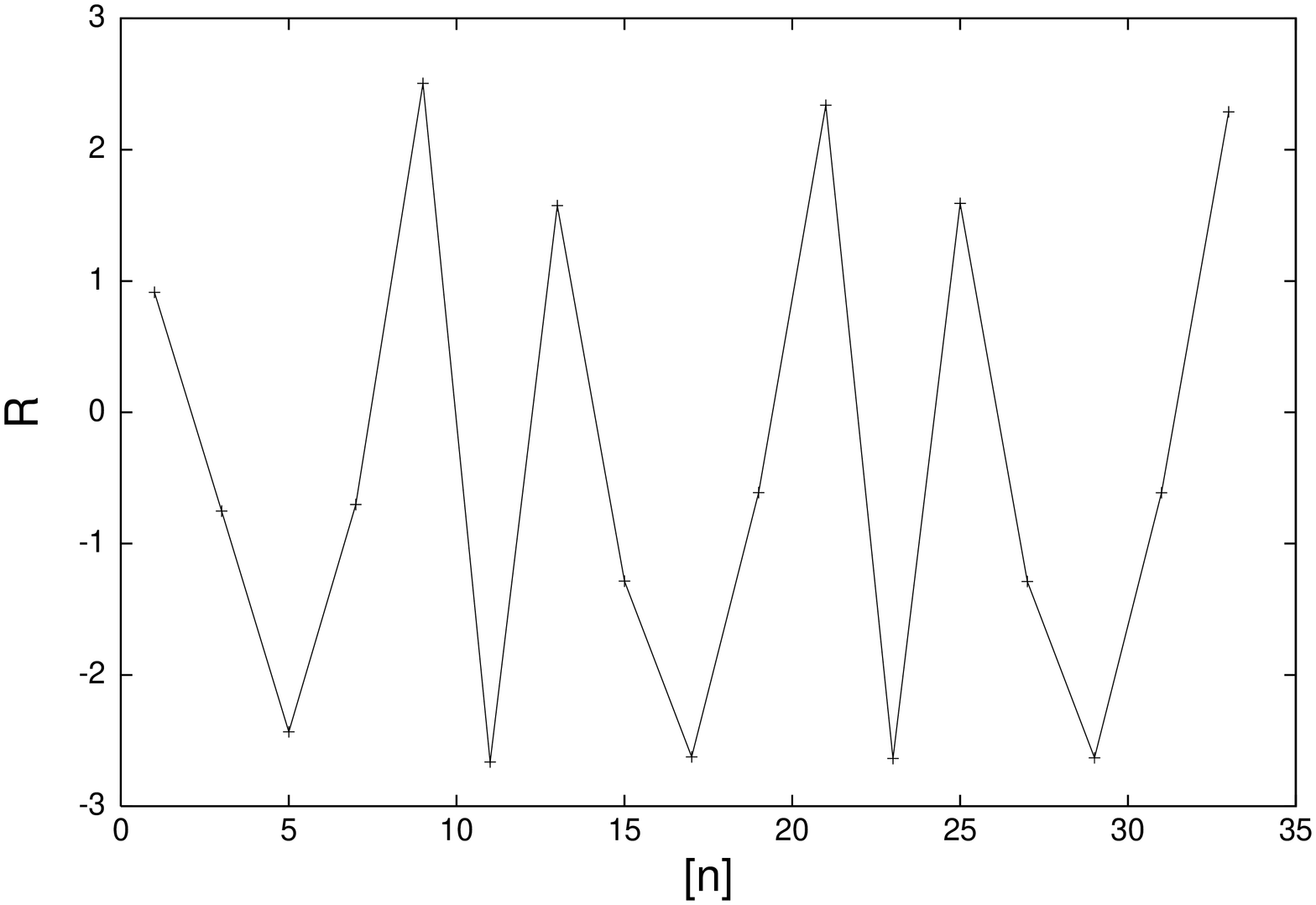}
    }
    \isubfig{			       
	 \includegraphics[angle=270,scale=2,width=.45\textwidth]{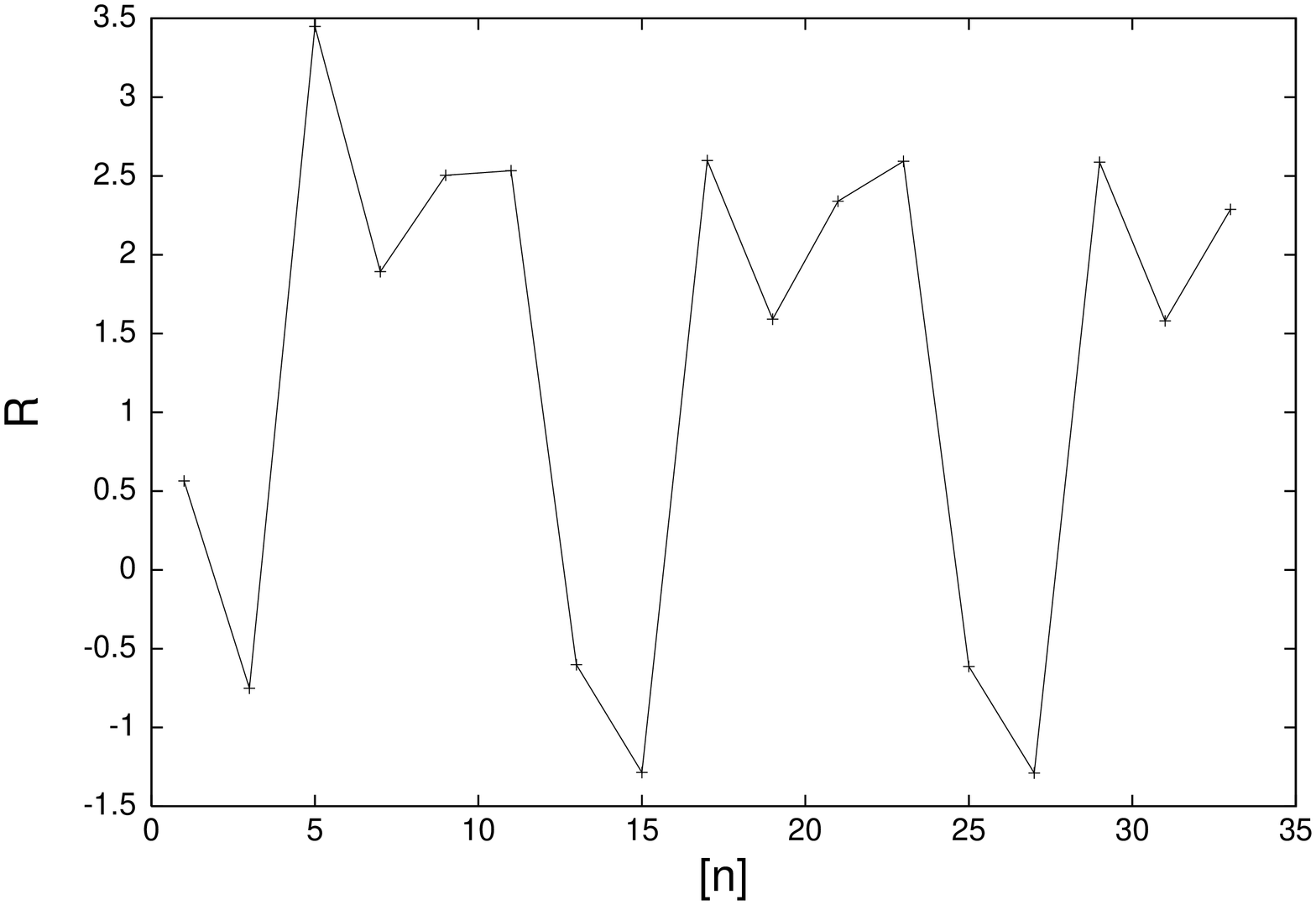}
    }
\caption{Residue behavior of the up (left figure) and down (right
  figure) periodic orbits on $s_1$ at the critical point on
  $1/\gamma^2$-bifurcation curve.}
\label{fig:updownorb}
\end{figure}

Finally, we found the critical point $(a_c,b_c)$ along the
$1/\gamma^2$-bifurcation curve to be the following:
\beq
a_c= 0.425160543   \qquad,\qquad b_c= 0.9244636470355.
\eeq
At the critical parameter values $(a_c,b_c)$, the residues of the down
periodic orbits on $s_1$, which are equal to the residues of the up
periodic orbits on $s_2$ because of the symmetry of the map (see
Eq.~(\ref{eq:tmap})), converge to the six-cycle\cite{note3} 
$\left\{ C_1, C_2, C_3, C_4, C_5, C_6\right\}$, where
\beq
\ba{ll} C_1 = -0.609\pm 0.005, & C_2= -1.288\pm 0.002,\\
C_3= 2.593\pm 0.005, & C_4= 1.584\pm 0.008,\\
C_5= 2.336\pm 0.006, & C_6= 2.593\pm 0.005.
\ea
\label{eq:cnum}
\eeq
The six-cycle can clearly be seen in Fig.~\ref{fig:updownorb} which
shows the residues of the up and down periodic orbits at the critical
point along the $s_1$ symmetry line. 
The residues of the up periodic orbits on $s_1$ (and of the down periodic
orbits on $s_2$) converge to the six-cycle
$\left\{ D_1, D_2, D_3,D_4, D_5, D_6\right\}$, where
\beq
\ba{ll} D_1 = 1.584\pm 0.008, & D_2= -1.288\pm 0.002,\\
D_3= -2.630 \pm 0.006, & D_4= -0.609\pm 0.005,\\
D_5= 2.336 \pm 0.006, & D_6= -2.630 \pm 0.006.
\ea
\label{eq:bnum}
\eeq
The residue convergence for other symmetry lines is shown in Table
\ref{tab:respat} where we denote by $R_{u_i}$ and $R_{d_i}$ the
residues of the up and down periodic orbit on the symmetry line $s_i$.
\begin{table}[ht]
\begin{center}
\begin{tabular}{|r|cccc|}\hline
&&&&\\[-2mm]
$[i]\quad\qquad$ & $R_{u_1}=R_{d_2}$ & $R_{u_2} = R_{d_1}$ &
$R_{u_3}=R_{d_4}$ & $R_{u_4} = R_{d_3}$\\[2mm]
\hline
&&&&\\[-2mm]
$ [1]\textrm{\ ,\ }[13]\textrm{\ ,\ }[25]$  & $D_1$ & $C_1$ & $D_4$ & $C_4$\\[1mm]
$ [3]\textrm{\ ,\ }[15]\textrm{\ ,\ }[27]$  & $D_2$ & $C_2$ & $D_5$ & $C_5$\\[1mm]
$ [5]\textrm{\ ,\ }[17]\textrm{\ ,\ }[29]$  & $D_3$ & $C_3$ & $D_6$ & $C_6$\\[1mm]
$ [7]\textrm{\ ,\ }[19]\textrm{\ ,\ }[31]$  & $D_4$ & $C_4$ & $D_1$ & $C_1$\\[1mm]
$ [9]\textrm{\ ,\ }[21]\textrm{\ ,\ }[33]$  & $D_5$ & $C_5$ & $D_2$ & $C_2$\\[1mm]
$[11]\textrm{\ ,\ }[23]\textrm{\ ,\ }[35]$  & $D_6$ & $C_6$ & $D_3$ & $C_3$\\[1mm]
\hline
\end{tabular}
\caption{\label{tab:respat} Period-six convergence pattern of the
residues near criticality along the different symmetry lines.} 
\end{center}
\end{table}
Note that the six-cycle $\{D_i\}$ of $R_{u_1}$ and $R_{d_2}$ (respectively, the
six-cycle $\{C_i\}$ of $R_{u_2}$ and $R_{d_1}$) is observed to be the same as
that of $R_{u_3}$ and $R_{d_4}$ (respectively, $R_{u_4}$ and
$R_{d_3}$) except it is shifted. The two six-cycles are related
because of the symmetry of the map as follows:
$D_1=C_4$, $D_2=C_2$, $D_4=C_1$, $D_5=C_5$, $C_3 =C_6$, and $D_3=D_6$.
It was numerically observed that $C_6\approx -D_6$, and therefore
$C_3\approx -D_3$. Using these relations we see that there are only five
independent residues which we take to be $C_1$,$C_2$,$C_3$,$C_4$, and $C_5$.

We compared the values of the residues at three different
points along the $1/\gamma^2$-bifurcation curve, one point
below criticality, one at criticality, and one above 
criticality: $(a_-,b_-)=(0.425160540,0.9244636195728)$,
$(a_c,b_c)=(0.425160543,0.9244636470355)$ and
$(a_+,b_+)=(0.425160545,0.9244636653440)$, respectively.
The numerical results for the $C_i$ are listed in Table \ref{tab:resnum1}.
\begin{table}[ht]
\begin{center}
\begin{tabular}{ | c || l | rrr | l | rrr | }\hline       
{[n]}  &       &  $z_-$ &  $z_c$  &  $z_+$ &       & $z_-$ & $z_c$ & $z_+$ \\
\hline
{[01]} & $C_1$ &  0.565 &  0.565  &  0.565 & $C_4$ & 0.914 & 0.914 & 0.914 \\
{[07]} &       & -0.702 & -0.702  & -0.702 &       & 1.893 & 1.893 & 1.893 \\
{[13]} &       & -0.601 & -0.601  & -0.601 &       & 1.574 & 1.574 & 1.574 \\
{[19]} &       & -0.611 & -0.611  & -0.612 &       & 1.590 & 1.591 & 1.592 \\
{[25]} &       & -0.610 & -0.612  & -0.614 &       & 1.578 & 1.591 & 1.599 \\
{[31]} &       & -0.566 & -0.612  & -0.644 &       & 1.406 & 1.581 & 1.710 \\
\hline
{[03]} & $C_2$ & -0.752 & -0.752  & -0.752 & $C_5$ & 2.169 & 2.169 & 2.169 \\
{[09]} &       & -1.328 & -1.328  & -1.328 &       & 2.505 & 2.505 & 2.505 \\
{[15]} &       & -1.286 & -1.286  & -1.286 &       & 2.329 & 2.329 & 2.329 \\
{[21]} &       & -1.289 & -1.290  & -1.291 &       & 2.337 & 2.340 & 2.341 \\
{[27]} &       & -1.273 & -1.289  & -1.300 &       & 2.300 & 2.338 & 2.364 \\
{[33]} &       & -1.161 & -1.249  & -1.276 &       & 1.873 & 2.288 & 2.614 \\
\hline
{[05]} & $C_3$ &  3.450 &  3.450  &  3.450 & $C_6$ & 3.450 & 3.450 & 3.450 \\
{[11]} &       &  2.534 &  2.534  &  2.534 &       & 2.534 & 2.534 & 2.534 \\
{[17]} &       &  2.598 &  2.598  &  2.598 &       & 2.598 & 2.598 & 2.598 \\
{[23]} &       &  2.588 &  2.594  &  2.598 &       & 2.588 & 2.594 & 2.598 \\
{[29]} &       &  2.498 &  2.588  &  2.650 &       & 2.498 & 2.588 & 2.650 \\
\hline
\end{tabular}
\caption{Numerical values of the residue six-cycle $C_i$
at $z_-=(a_-,b_-)$, $z_c=(a_c,b_c)$, and $z_+=(a_+,b_+)$.}
{\label{tab:resnum1}}
\end{center}
\end{table}
We see that each element of the six-cycle tends to
zero for $\left(a_-,b_-\right)$, to infinity for
$\left(a_+,b_+\right)$, while it tends to the critical value at
$\left(a_c,b_c\right)$. 
\begin{figure}[h!t]
    \centering
    \isubfig{
	 \includegraphics[angle=270,scale=2,width=.45\textwidth]{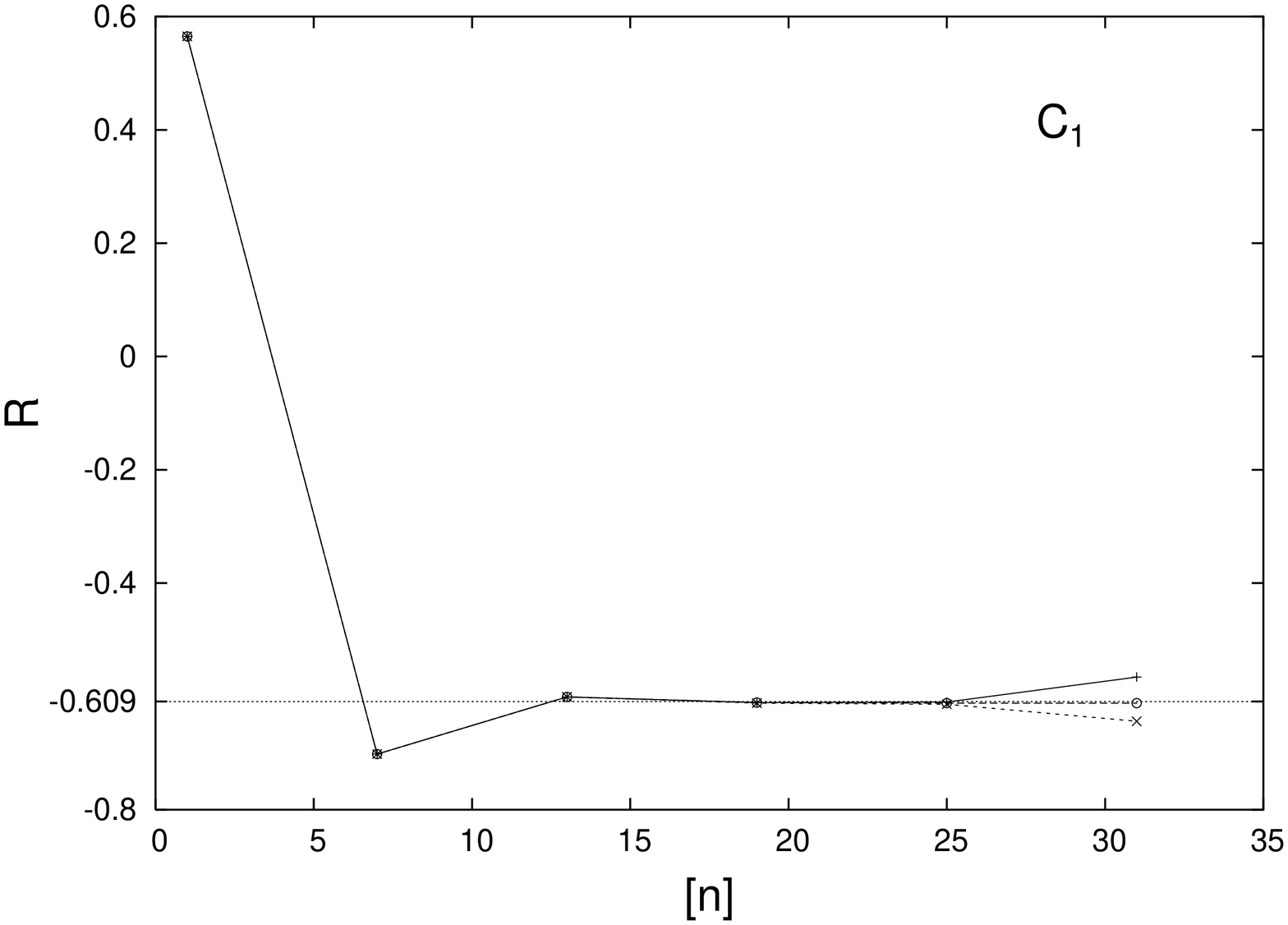}
    }
    \isubfig{			       
	 \includegraphics[angle=270,scale=2,width=.45\textwidth]{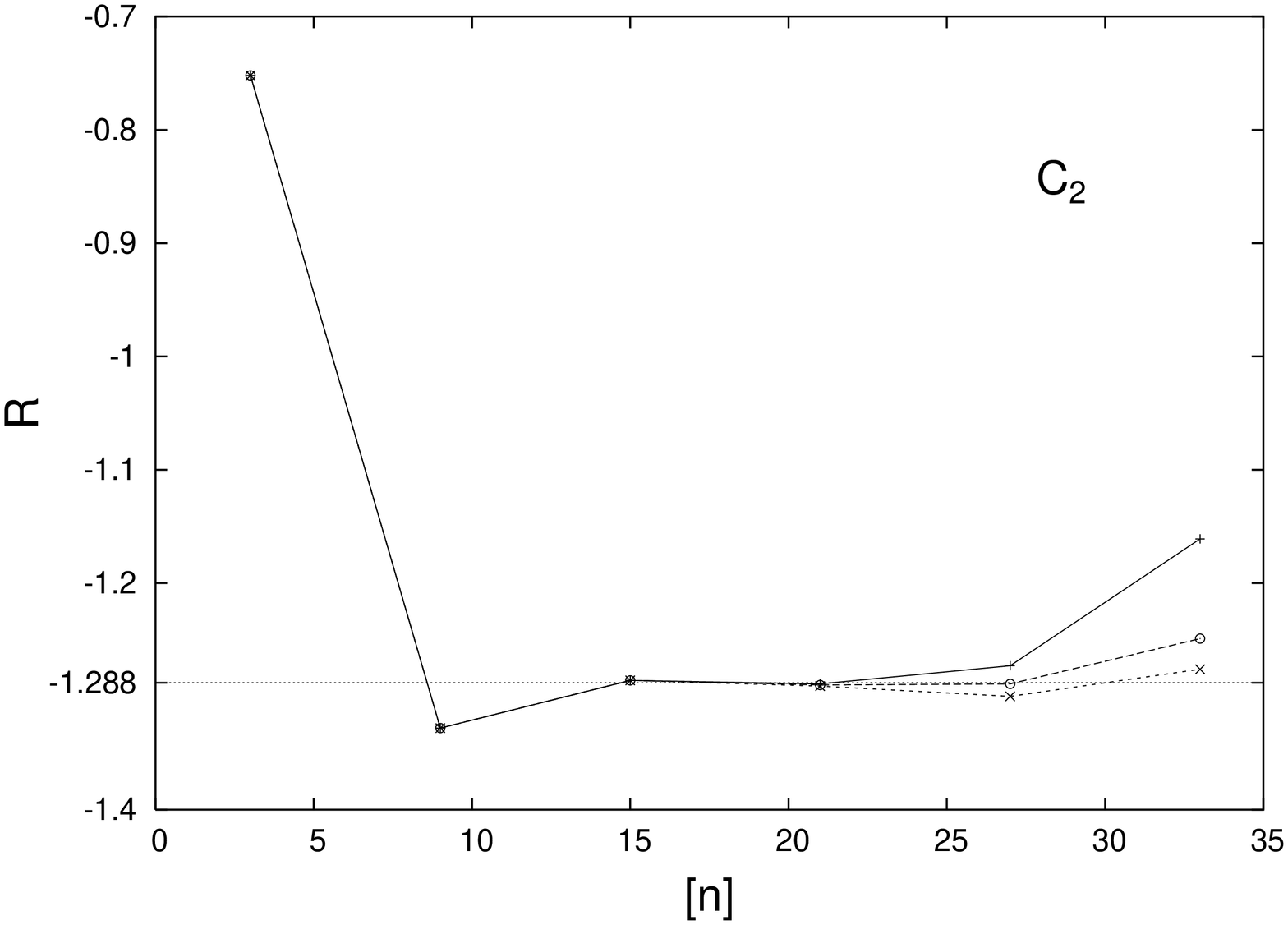}
    }
    \isubfig{			       
	 \includegraphics[angle=270,scale=2,width=.45\textwidth]{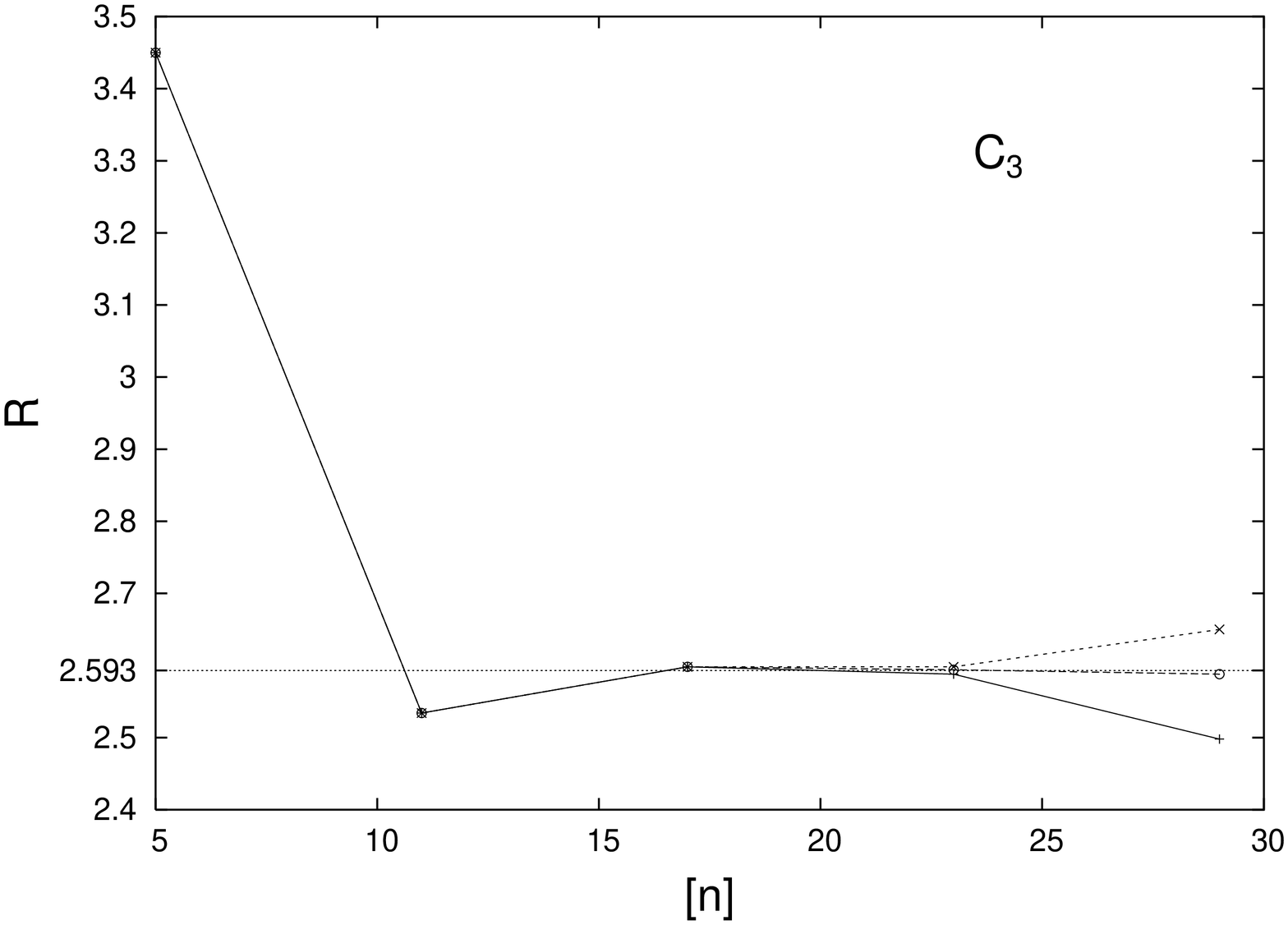}
    }
    \isubfig{			       
	 \includegraphics[angle=270,scale=2,width=.45\textwidth]{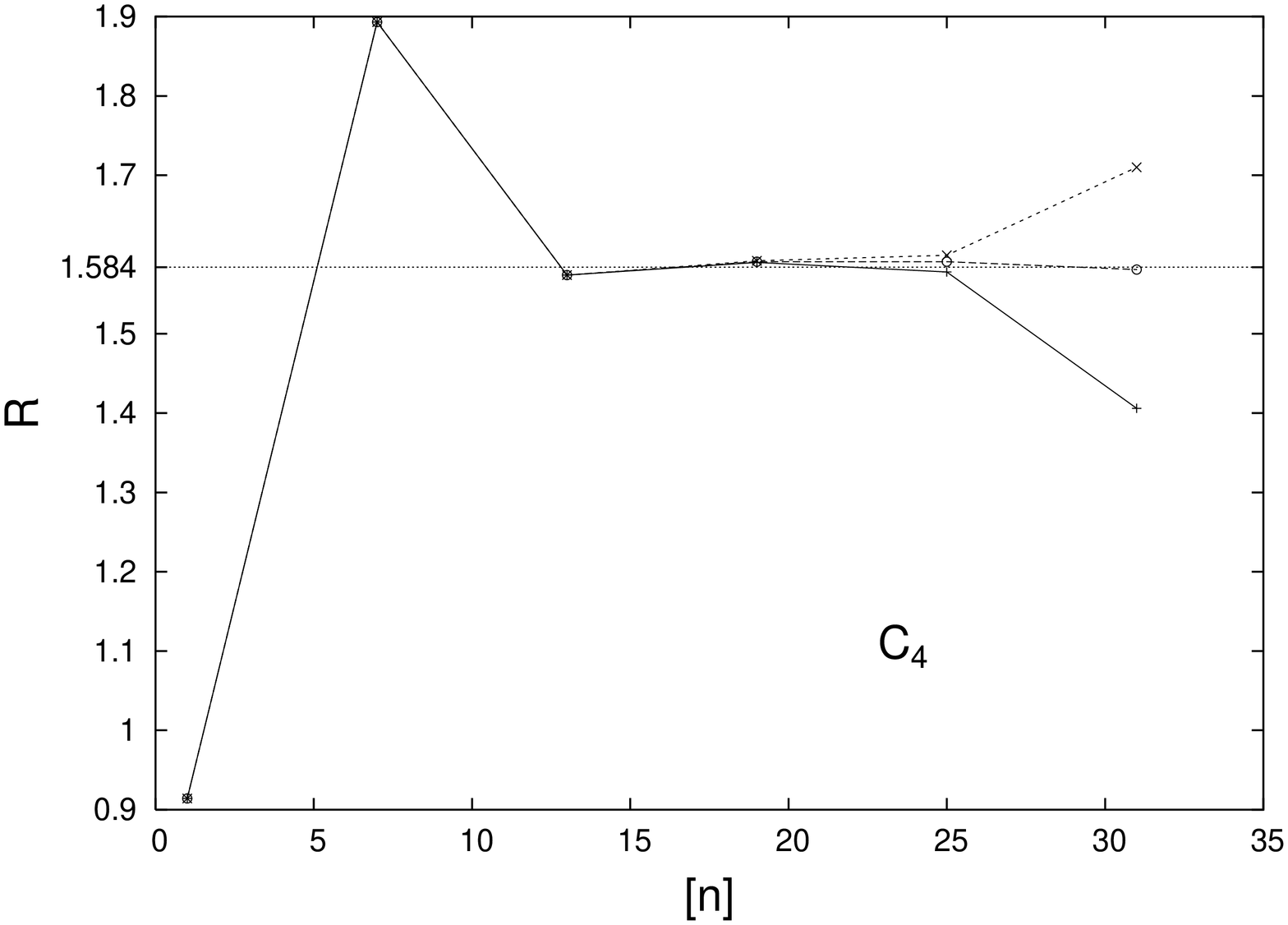}
    }
    \isubfig{			       
	 \includegraphics[angle=270,scale=2,width=.45\textwidth]{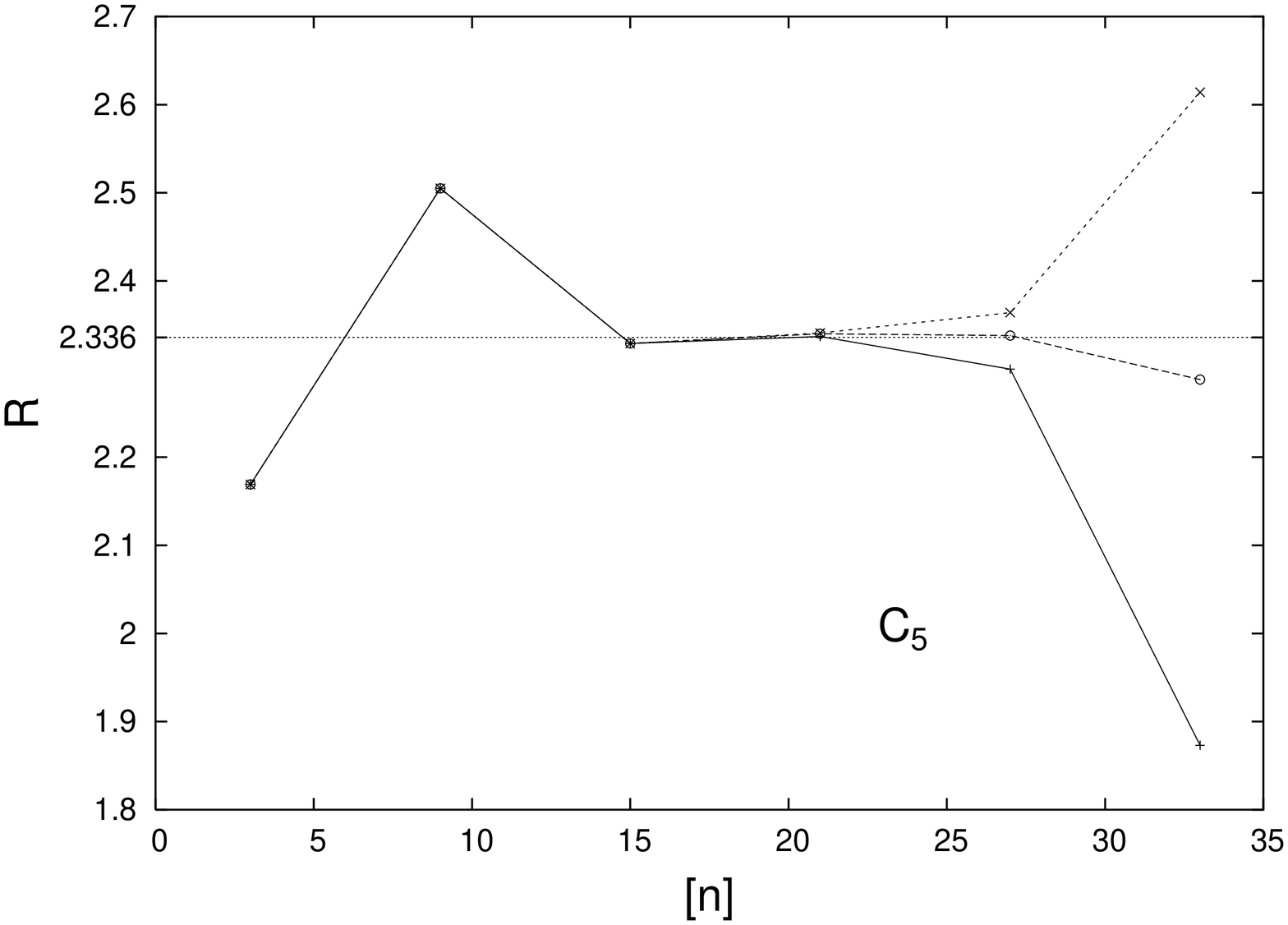}
    }
    \isubfig{			       
	 \includegraphics[angle=270,scale=2,width=.45\textwidth]{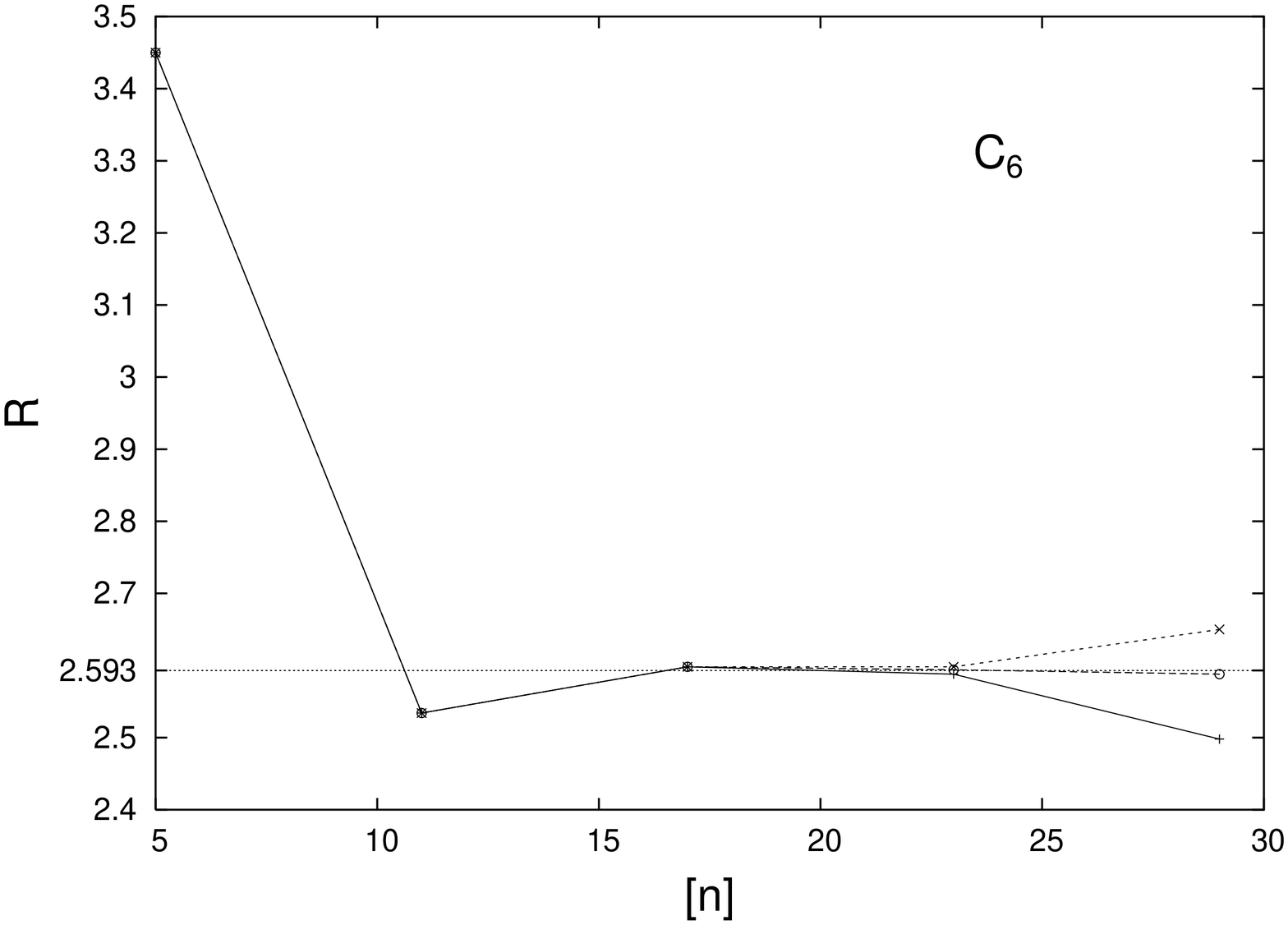}
    }
\caption{Residue Convergence for $C_1$ through $C_6$ at $z_1$ ($+$),
$z_2$ ($\circ$) and $z_3$ ($\times$) (see Table \ref{tab:resnum1}). }
\label{fig:res}
\end{figure}
Figure \ref{fig:res} clearly illustrates this behavior.

A comparison with the results of Ref.~\onlinecite{diego4}
shows that, within numerical accuracy, we found the same values for
the residues of the six-cycle, but the sequence has shifted by two:
$C_1 = H_3$, $C_2= H_4$, $C_3=H_5$, $C_4= H_6$, $C_5= H_1$ and $C_6=
H_2$, where $H_i$ denote the residues for the $1/\gamma$ case found in
Ref.~\onlinecite{diego4}. A similar shift by two occurs for other
symmetry lines.

\subsubsection{Spatial scaling at criticality}

\begin{figure}[h!t]
\centering
    \isubfig{
	 \includegraphics[angle=270,scale=2,width=.45\textwidth]{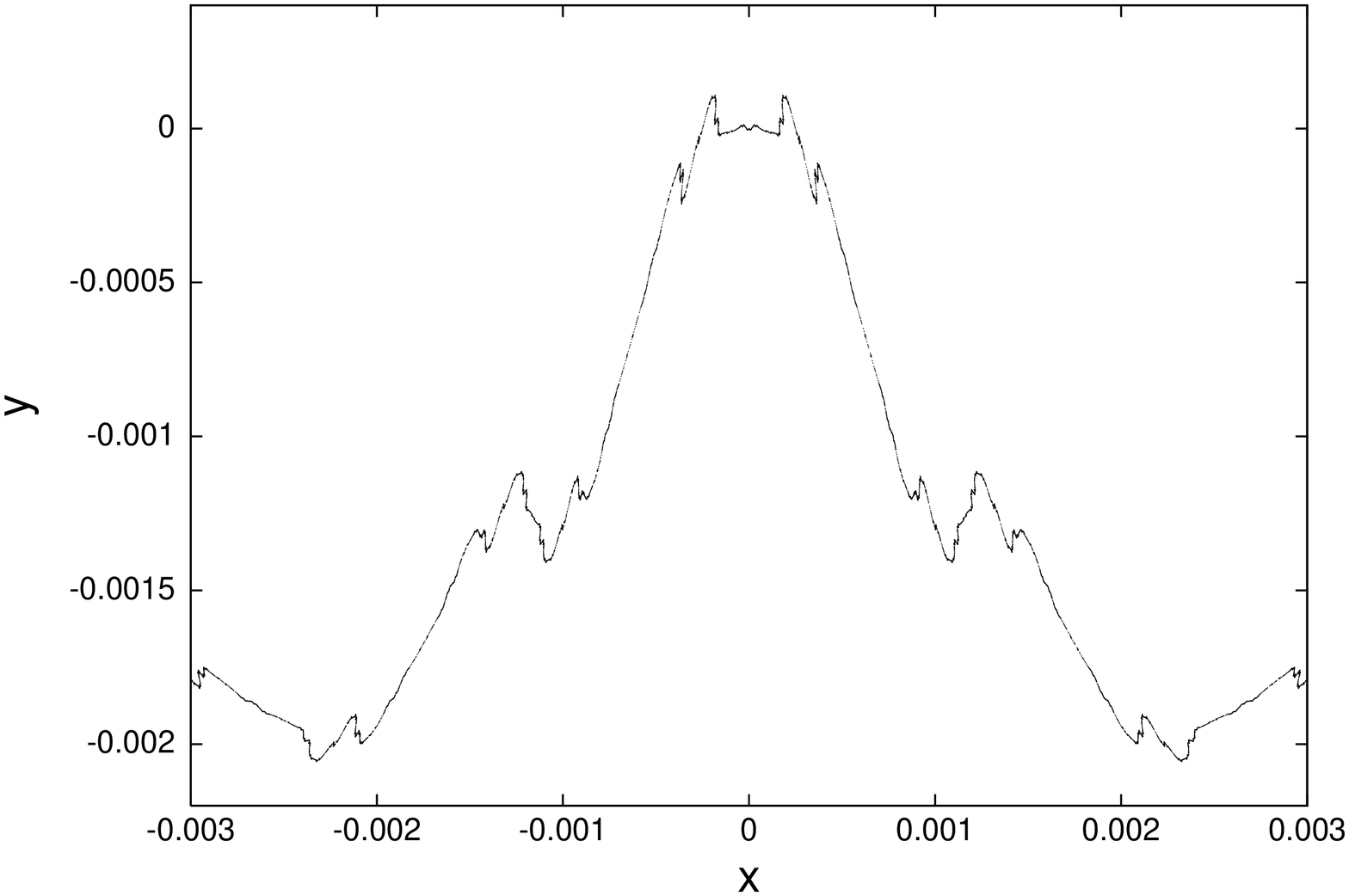}
    }
    \isubfig{			       
	 \includegraphics[angle=270,scale=2,width=.45\textwidth]{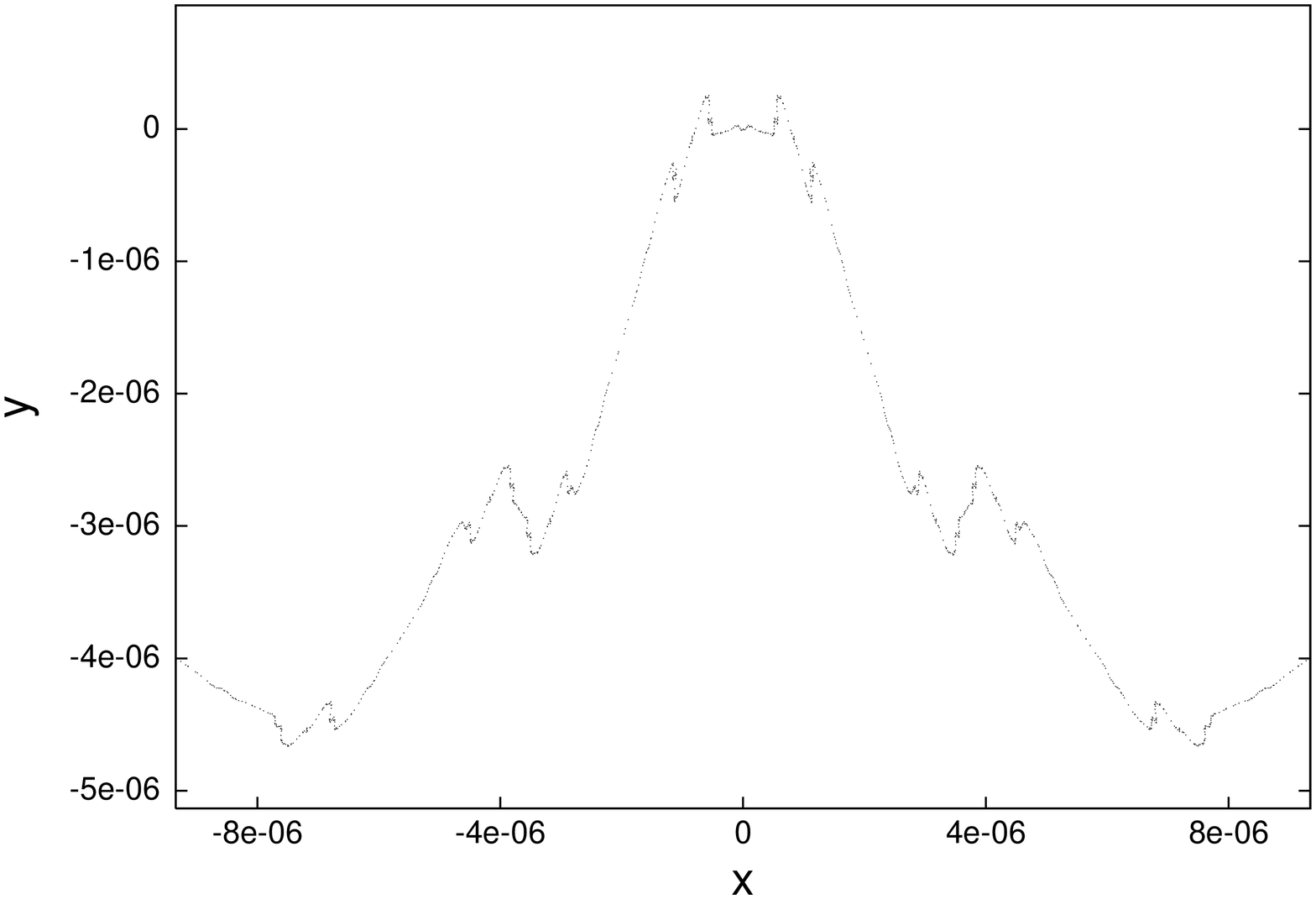}
    }
\caption{Invariance under rescaling of shearless $1/\gamma^2$ torus at
criticality.} \label{fig:torusatcrit}
\end{figure}

As expected, the shearless curve exhibits scale invariance at
criticality, which can be demonstrated explicitly  by using {\it
symmetry line coordinates}\cite{diego4} $(\hat{x},\hat{y})$ defined by
$\hat{x}=x- a( 1-y^2)/2$ and $\hat{y}=y-y_s$.
In these coordinates, the $s_3$ symmetry line becomes a straight line
that intersects the shearless curve at the origin. We find that, in
symmetry line coordinates, the shearless $1/\gamma^2$ invariant
torus at criticality remains invariant under a scale change $(x,y) \to
(\alpha^{12} x, \beta^{12} y)$. This property is illustrated in
Fig.~\ref{fig:torusatcrit}.

As described in Ref.~\onlinecite{diego5}, we can find $y_s$ using
\beq
y_s = \lim_{i\to\infty} \frac{y_{[2i+1]}\; y_{[2i+11]} - y_{[2i-1]}\; 
y_{[2i+13]}}{\left(y_{[2i+1]}-y_{[2i-1]}\right) - \left(
y_{[2i+13]}-y_{[2i+11]}\right)} \approx 0.47253494777 ,
\eeq
\noindent
where $y_{[n]}$ denotes the $y$-coordinate of the periodic orbit
$[n]$ along the $s_3$ symmetry line. To obtain the quoted value of
$y_s$ we used $i=10$. We then obtained $\alpha$ and $\beta$ as
follows:\cite{note4}
\beq
\alpha = \lim_{n\to\infty} \left| 
\frac{\hat{x}_{[2i+1]}}{\hat{x}_{[2i+13]}}\right|^{1/12} \approx 1.61759
\label{eq:alphaus}
\eeq
and
\beq
\beta= \lim_{n\to\infty} \left| 
\frac{\hat{y}_{[2i+1]}}{\hat{y}_{[2i+13]}}\right|^{1/12} \approx 1.65702,
\label{eq:betaus}
\eeq
where $\left( \hat{x}_{[n]},\hat{y}_{[n]}\right)$ are symmetry line
coordinates of the point of the periodic orbit $[n]$ that is
the closest to the origin. Within numerical accuracy, these values are
the same as in Ref.~\onlinecite{diego5}.

\begin{figure}[ht]
\centering
\includegraphics[angle=270,scale=2,width=0.9\textwidth]{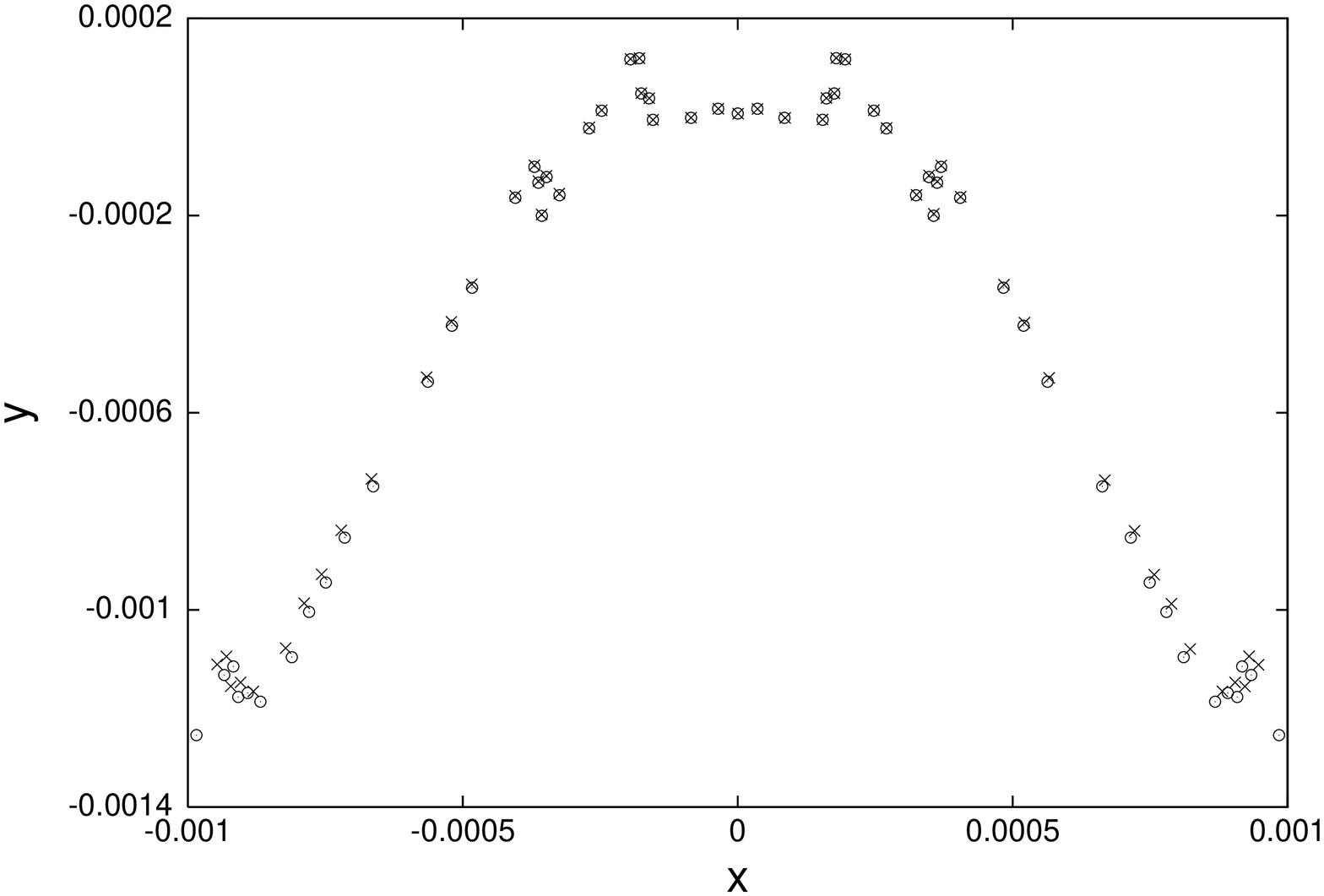}
\caption{Invariance of periodic orbits under simultaneous rescaling
  and shift of winding numbers by twelve. Here we show the periodic
  orbits $[21]$ ($\times$) and $[33]$ ($\circ$) after rescaling of $x$
  and $y$ coordinates by $\alpha^{12}$ and $\beta^{12}$ respectively.}
\label{fig:periodscale}
\end{figure}

Further numerical analysis shows that periodic orbits also exhibit
scaling behavior locally near the $s_3$ symmetry line.
Figure \ref{fig:periodscale} shows points of the periodic
orbit $[n]=[21]$ (in symmetry line coordinates) and points of the
periodic orbit $[33]$ with the $x$ and $y$ coordinates rescaled by
$\alpha^{12}$ and $\beta^{12}$ respectively. The result suggests that
periodic orbits remain invariant under a simultaneous spatial
rescaling and shifting of the winding number by twelve from $[n]$ to
$[n+12]$.

\subsection{Numerical accuracy}\label{sssec:erroran}

We conclude this section with comments about the numerical accuracy of
the results.

\begin{enumerate}
\item Points on the $m/n$-bifurcation curves were found
with an accuracy ranging between $10^{-12}$ and $10^{-15}$,
where the larger value corresponds to larger periods.
We obtain this measure of accuracy from the condition
\beq
F(y) =0,\quad F'(y)=0\quad \mbox{ and }\quad F''(y)\ne 0 ,
\eeq
as explained previously. The numbers quoted above are 
the values of $F(y)$ obtained at the numerically found
minima in $y$.
\item Periodic orbits along the different symmetry lines
around the critical point where found with an accuracy
ranging between $10^{-15}$ and $10^{-17}$. Here, the criterion
is the difference between the winding number $m/n$ of
the periodic orbit of interest, and the winding number of
the orbit that results when
starting at the numerically found location of the periodic
orbit on the respective symmetry line, and then
iterating the map.
\item A criterion for the accuracy of the scaled bifurcation curve
$\Phi_{1/\gamma^2}$ is harder to find, since the location of
the actual curve is unknown. An upper bound
on the error, though, should be the distance between the
$\Phi_{1/\gamma^2}$ and $\Phi_{[37]}$, since the latter 
definitely lies on the other side of $\Phi_{1/\gamma^2}$. This error was
found to be approximately $2\times10^{-13}$.
\item A criterion for the accuracy of the critical point
in parameter space, $(a_c,b_c)$ is even harder to
define, since we cannot actually find the residues of existing orbits
of {\it all } periods, which is required to check if the
six-cycle at that point continues {\it ad infinitum}. We believe that
the value for  $a_c$ is accurate up to $1\times 10^{-9}$.
\item The uncertainties for the critical residues, quoted above in
Eq.~(\ref{eq:cnum}), were computed from the variation in numerical values
of residues at $a_c$ for the three or four highest period orbits found (\eg
for $C_1$, using residues of $[13]$, $[19]$, $[25]$ and $[31]$).
\item If we evaluate the residues of the up and down periodic orbits along
the $s_1$ and $s_3$ symmetry lines, then the residues on the other two
symmetry lines can be constructed using symmetry arguments (see
Sec.~\ref{ssec:po}). But, as a check of the numerical procedures,
we independently evaluated the residues on all the four symmetry lines
and confirmed the symmetry arguments.
\end{enumerate}

\section{Renormalization group interpretation}\label{sec:reng}

In this section, we interpret the above results within the 
renormalization group framework. The analysis follows Refs.~\onlinecite{diego5}
closely, since, as expected, the residue behavior exhibits a six-cycle
at criticality. But, because of the different winding number (\ie
different from $1/\gamma$) the renormalization group operator
will have a different form.

Renormalization ideas have been used fruitfully in area-preserving
maps and Hamiltonian flows. (See \eg Refs.~\onlinecite{shenker,mackay1,
greene4,escande,chandre3,abad,koch} and references
therein.) In contrast to mathematical KAM theory, which proves the
existence of dense sets  of invariant tori in regions of phase space,
the renormalization group approach addresses the problem of the
destruction of an invariant torus with a {\it specific} winding number
under strong perturbation.

The following renormalization approach (see \eg
Refs.~\onlinecite{greene6,diego5}) is based on the residue criterion
(Sec.~\ref{ssec:po}). To study the breakup of an invariant torus of
winding number $\omega$, we loosely represent the map $M$ as:
\beq
M = \left( R_1, R_2, R_3, \ldots\right) ,
\label{eq:lineofres}
\eeq
where the $\{R_i\}$ are the residues of the periodic orbits with
winding numbers $\{m_i/n_i\}$, the convergents of $\omega$. For
example, the integrable map will be represented by
$\left(0,0,0,\ldots\right)$ because all the orbits are parabolic in
that case. The key idea is to construct an operator ${\mathcal R}$
that explores the infinite tail of  (\ref{eq:lineofres}) by mapping a
map given by (\ref{eq:lineofres}) to another map, ${\mathcal R}(M)$,
represented by ${\mathcal R}\left(M\right)= \left(
\hat{R}_1,\hat{R}_2,\ldots, \hat{R}_i,\ldots\right)$,
where $\hat{R}_i=R_{i+1}$. This operation can  be interpreted as a
{\it time renormalization} since periodic orbits with large periods
are transformed into periodic orbits with smaller periods, which
amounts to a rescaling of time.

The residue criterion can now be rephrased in this framework:
\begin{enumerate}
\item If $\lim_{n\to\infty} {\mathcal R}^n(M) =
 \left( 0,0,0,\ldots\right)$, the invariant torus exists.
\item If $\lim_{n\to\infty} {\mathcal R}^n(M) =\left(\pm\infty, 
\pm\infty,\ldots\right)$, the invariant torus is destroyed.
\item If $\lim_{n\to\infty} {\mathcal R}^n(M)$ is a map for
which the residues have finite, nonzero values, \ie a
map that is invariant under the action of ${\mathcal R}^m$ (a fixed
point of ${\mathcal R}^m$) for some $m>0$, the
invariant torus is at the threshold of destruction. Possible
scenarios are the convergence of the residues to a fixed value
or to a convergent subsequence.
\end{enumerate}
There are two kinds of fixed points: {\it simple} fixed points and
{\it critical} fixed points. In the case of area-preserving maps, we
come to the following interpretation.
A {\it simple} fixed point is an integrable map (all the residues are
zero), and its basin of attraction contains all the maps for which the
invariant torus exists.
A {\it critical} fixed point is a map for which the invariant torus
under consideration is at criticality. All the maps in its basin of
attraction exhibit the same universal behavior at the critical breakup.

\subsection{Renormalization group operator}\label{sssec:operator}

Following the discussion in Refs.~\onlinecite{mackay1} and
\onlinecite{diego5}, we use pairs of commuting maps because they
provide a simple way to define the renormalization operators for
invariant tori. 

A {\it pair of commuting maps} is an ordered pair of maps,
$(U,T)$, such that $UT=TU$. An {\it orbit} of a point $(x,y)$
generated by $(U,T)$ is the set of points $\{U^m T^n (x,y)\}$,
where $m$ and $n$ are integers. A {\it periodic orbit of period $m/n$}
is an orbit for which $U^{m}T^{n}(x_i,y_i) = (x_i,y_i)$.

For the breakup of the invariant torus with winding number
$\omega=1/\gamma^2$, we define the renormalization group operator by
\beq
{\mathcal R} \left(\ba{c} U \\ T\ea\right) := B \left(\ba{c} U^{-1} T^{-1}\\
U\; T^2\ea\right) B^{-1}.\label{eq:rgodef}
\eeq

\noindent
As for the case of $1/\gamma$,\cite{diego5} this operator contains both
time and space renormalization as follows: 

The {\it space renormalization} is represented by the operator $B$,
which rescales the $(x,y)$ coordinates, \ie $(x,y)\to B(x,y)$ where
\beq
B = \left(\ba{cc} r & 0\\ 0 & s \ea\right).
\eeq
At the critical fixed point studied in this paper, we see that
$r=\alpha$ and $s=\beta$ given by Eqs.~(\ref{eq:alphaus})-(\ref{eq:betaus}).

The {\it time renormalization} is, again, accomplished by the specific
combination of the commuting maps.
If $(x,y)$ is a periodic orbit of $(U,T)$ with winding number
$F_{i}/F_{i+2}$, then $B (x,y)$ is a periodic orbit of
$(\hat{U},\hat{T}) = {\mathcal R}\left(U,T\right)$
with winding number $F_{i-1}/F_{i+1}$, as can be verified as follows:
\bean
\hat{U}^{F_{i-1}}\;\hat{T}^{F_{i+1}} B (x,y)
&=& B (U T)^{-F_{i-1}}\; \left(U T^2\right)^{F_{i+1}} (x,y)\\
&=& B U^{-F_{i-1}+F_{i+1}}\; T^{-F_{i-1}+2F_{i+1}} (x,y)\\
&=& B U^{F_i}\; T^{F_{i+2}} (x,y)\\
&=& B (x,y).
\eean
By induction, an orbit with winding number $F_{i}/F_{i+2}$ under
$(U,T)$ is transformed into an orbit of ${\mathcal R}^n(U,T)$ with
winding number $F_{i-n}/F_{i+2-n}$.

\subsection{Simple periodic orbit of ${\mathcal R}$}

We can find the integrable period-two orbit $(U_\pm,T_\pm)$ of
the renormalization operator (\ref{eq:rgodef}) by requiring that 
${\mathcal R}(U_\pm,T_\pm)=(U_\mp,T_\mp)$. This two-cycle is given by
the following pairs of maps:
\beq
U_\pm\xyvect{x}{y} = \xyvect{x-\gamma^2\mp y^2/\gamma^2}{y}, \qquad
T_\pm\xyvect{x}{y} = \xyvect{x+1 \pm y^2}{y},
\eeq
where the rescaling of the coordinates is given by
\beq
B = \left(\ba{cc} -\gamma & 0\\ 0 & \pm\gamma \ea\right).
\eeq

Using the definition $U_\pm^m T_\pm^n(x,y)=(x,y)$ of the periodic
orbits of period $m/n$, we get the rotation number as a function of $y$:
\bean
\omega_\pm(y) &=& -\frac{1\pm y^2}{-\gamma^2 \mp y^2/\gamma^2}
             \;=\; \frac{1}{\gamma^2}\left(1\pm y^2\right)
                \left(1\pm\frac{y^2}{\gamma^2}\right)^{-1}\\
       &\approx& \frac{1}{\gamma^2}\left[1\pm
                \left(1-\frac{1}{\gamma^4}\right)y^2 + \ldots \right].
\eean
Thus we see that the map $(U_-,T_-)$ is
locally equivalent, under a change of coordinates, to the SNM with
parameters $(a,b)=(1/\gamma^2,0)$.

\subsection{Critical periodic orbit of ${\mathcal R}$}

The next step is to analyze the critical periodic orbit of ${\mathcal R}$.
 Consider the nontwist map
\beq
{\mathcal C} = \left( C_1,-,C_2,-,C_3,-,C_4,-,C_5,-,C_6,-,C_1,-,C_2,
\ldots\right),
\eeq
where the $C_i$ are the elements of the six-cycle computed earlier, 
and the ``-'' denote the periodic orbits that do not exist (see Table
 \ref{tab:conv2}). By construction,
this map is a period-12 orbit of the renormalization group
operator (a fixed point of ${\mathcal R}^{12}$), \ie
\beq
{\mathcal R}^{12}{\mathcal C} ={\mathcal C}.\label{eq:fixpt}
\eeq
In Sec.~\ref{ssec:results}, we found that the residues of the
periodic orbits approximating the $1/\gamma^2$-shearless curve in the
standard nontwist map exhibit
convergence to 
the six-cycle $\left\{ C_i\right\}$.
Assuming that we can fine-tune the results for 
$\left(a_c,b_c\right)$, we expect that
$\lim_{n\to\infty} {\mathcal R}^n M \left(a_c,b_c\right) = {\mathcal C}$.

If we are studying the breakup of the $1/\gamma^2$-shearless curve for
parameter values along the bifurcation curve for one of the low-period
convergents, then we start near the
stable manifold of the critical periodic orbit of ${\mathcal R}$. But,
under the action of ${\mathcal R}$, we are pushed along an
unstable direction. Thus, we see parts of the six-cycle of residues
(see Fig.~\ref{fig:updownorb2}), but the limiting residue behavior
is observed to be $\lim_{i\to\infty} \left|R_i\right| \approx 0.25$,
which is characteristic for the critical fixed point of {\it twist}
maps (see \eg Ref.~\onlinecite{mackay1}). In renormalization group
language, this means that part of the unstable manifold of the
critical nontwist fixed point (maps for which $(a,b)$ is below
$(a_c,b_c)$) is in the basin of attraction of the critical twist fixed
point. 

\subsection{Eigenvalues}

As shown in Ref.~\onlinecite{diego5}, it is possible to use our numerical
data to draw further conclusions about the renormalization group
operator ${\mathcal R}$, in particular to compute its unstable
eigenvalues. 
The main difficulty in computing these eigenvalues is that the space
of maps is infinite-dimensional whereas the $(a,b)$ parameter space
has only two dimensions.
But the fact that we can find an isolated point $\left(a_c,b_c\right)$
in parameter space at which the map is at
criticality means that the dimension of the
unstable manifold is two.
The map $M$ at $\left(a_c,b_c\right)$ is the intersection
point of the two-parameter family of maps with
the stable manifold of the fixed point, \ie
values $a_c$ and $b_c$ describe the location
of the critical fixed point of ${\mathcal R}^{12}$ in its unstable 
manifold.

Below, we first compute the eigenvalues
characterizing the approach to $\left(a_c,b_c\right)$
in the $(a,b)$ parameter space using the numerical
results from above. 
As shown in Ref.~\onlinecite{diego5}, based on the type of
numerical data obtained, the two eigenvalues can be found  by

\beq
\nu_1 = \lim_{n\to\infty} \left(
 \frac{\Phi_{[n+12]}\left(a_c\right)-b_c}{\Phi_{[n]}\left(a_c\right)-b_c}\right)
\label{eq:nu1def}
\eeq
and
\beq
\nu_2 = \lim_{n\to\infty} \left( \frac{a_{c\,[2n+12]}-a_c}{a_{c\,[2n]}-a_c}\right).
\label{eq:nu2def}
\eeq

The last step is to relate the values $\nu_i$ to the unstable
eigenvalues $\delta_i$ of the renormalization group operator
${\mathcal R}$.
The key idea is to study the behavior of the residues of the
periodic orbits approximating $\omega$ at the $(a,b)$ values
used in the computation of $\nu_i$. For details see Ref.~\onlinecite{diego5}.

Denoting the unstable eigenvalues of ${\mathcal R}$ by $\delta_1$
 and $\delta_2$, we conclude that
$\delta_i = \left(1/\nu_i\right)^{1/12}$.
We find numerical values of
\beq
\delta_1 \approx 2.678\,,\qquad \delta_2 \approx 1.583.
\eeq
Comparing this to the values found in Ref.~\onlinecite{diego5} shows that within
(assumed) numerical uncertainty these values are the same as those
for $1/\gamma$, as predicted.
The $a_c$ values used to determine $\delta_2$ were
$a_{c[26]}$ and $a_{c[14]}$, which explains the larger
discrepancy. Work is under way to improve this result.

\section{Conclusion}\label{sec:concl}

We have shown through numerical simulations that the critical
residue values at the breakup of the $1/\gamma^2$-shearless curve
in the standard nontwist map coincide with those of the
 $1/\gamma$-shearless curve. In addition, the critical scaling 
parameters and the unstable eigenvalues of the renormalization
group operator were found to be the same for both cases.
The main differences are the location of the respective critical
point in parameter space and the detailed form of the
renormalization group operator in terms of commuting maps pairs.

Future work includes the search for the breakup values of
more winding numbers to map out the details of the critical function
depicted in Fig.~\ref{fig:stntshin2}. In addition, new fixed points of the
renormalization group operator might be obtained by studying the breakup of
shearless curves with non-noble winding numbers.

\acknowledgments
The authors would like to thank John Greene and Diego del-Castillo-Negrete
for many helpful discussions.
This research was supported in part by U.S. Department of Energy
Contract No. DE-FG01-96ER-54346 and by an appointment of A.~Wurm
to the U.S. Department of Energy Fusion Energy Postdoctoral Research
Program administered by the Oak Ridge Institute for Science and 
Education.


\begin{thebibliography}{99}

\bibitem{diego2} D.~del-Castillo-Negrete and P.J.~Morrison,
``Chaotic transport by Rossby waves in shear flow,''
Phys.~Fluids A {\bf 5}, 948 (1993).

\bibitem{petrisor1} E.~Petrisor, ``Nontwist area preserving maps
with reversing symmetry group,'' Int.~J.~Bif.~Chaos {\bf 11},
497 (2001).

\bibitem{balescu} R.~Balescu, ``Hamiltonian nontwist
maps for magnetic field lines with locally reversed
shear in toroidal geometry,'' Phys.~Rev.~E {\bf 58},
3781 (1998).

\bibitem{horton1} W.~Horton, H.-B.~Park, J.-M.~Kwon,
D.~Strozzi, P.J.~Morrison and D.-I.~Choi, ``Drift wave
test particle transport in reversed shear profile,''
Phys.~Plasmas {\bf 5}, 3910 (1998).

\bibitem{kyner} W.T.~Kyner, ``Rigorous and formal stability
of orbits about an oblate planet,'' Mem.~Am.~Math.~Soc. {\bf 81},
1 (1968).

\bibitem{chandre2} C.~Chandre, D.~Farrelly and T.~Uzer, ``Threshold
to chaos and ionization for the hydrogen atom in rotating fields,''
Phy.~Rev.~A {\bf 65}, 053402 (2002).

\bibitem{dullin} H.R.~Dullin, J.D.~Meiss and D.~Sterling,
``Generic twistless bifurcations,'' Nonlinearity {\bf 13}, 202
(2000).

\bibitem{vander} J.P.~Van Der Weele, T.P.~Valkering, H.W.~Capel,
  T.~Post, ``The birth of twin Poincar\'e-Birkhoff chains near $1:3$
  resonance,'' Physica {\bf A 153}, 283 (1988); J.P.~Van Der Weele and
  T.P.~Valkering, ``The birth process of periodic orbits in non-twist
  maps,'' Physica {\bf A 169}, 42 (1990).

\bibitem{delshams} A.~Delshams and R.~de~la~Llave, ``KAM
theory and a partial justification of Greene's
criterion for non-twist maps,'' SIAM~J.~Math.~Anal. {\bf 31},
1235 (2000).

\bibitem{franks} J.~Franks and P.~Le~Calvez, ``Regions of instability
for nontwist maps,'' preprint, math.DS/9910152, Los Alamos
(1999).

\bibitem{simo} C.~Sim\'o, ``Invariant curves of analytic
perturbed nontwist area preserving maps,'' Regular and
Chaotic Dynamics {\bf 3}, 180 (1998).

\bibitem{diego4} D.~del-Castillo-Negrete, J.M. Greene and
P.J.~Morrison, ``Area preserving nontwist maps: periodic orbits
and transition to chaos,'' Physica~D {\bf 91}, 1 (1996).

\bibitem{diego5} D.~del-Castillo-Negrete, J.M. Greene and
P.J.~Morrison, ``Renormalization and transition to chaos in
area preserving nontwist maps,'' Physica~D {\bf 100}, 311
(1997).

\bibitem{shin1} S.~Shinohara and Y.~Aizawa, ``The Breakup
Condition of Shearless KAM Curves in the Quadratic Map,''
Progr.~Theo.~Phys. {\bf 97}, 379 (1997).

\bibitem{note1} Note, that the map used
in Ref.~\onlinecite{shin1} is different from the map we use
(Eq.~(\ref{eq:stntmap})), but they are related by a coordinate
transformation, including rescaling of parameters.

\bibitem{stark} J.~Stark, ``Determining he critical transition for
circles of arbitrary rotation number,'' Phys.~Lett. {\bf A 163}, 258
(1992).

\bibitem{greene2} J.M.~Greene, ``A method for computing the
stochastic transition,'' J. Math. Phys. {\bf 20}, 1183 
(1979).

\bibitem{khinchin} A.Ya.~Khinchin, {\it Continued Fractions},
3rd. ed., University of Chicago Press, Chicago, IL (1964).

\bibitem{falc} C.~Falcolini and R.~de~la~Llave, ``A rigorous
partial justification of Greene's residue criterion,'' 
J.~Stat.~Phys. {\bf 67}, 609 (1992).

\bibitem{mackay2} R.S.~MacKay, ``On Greene's residue
criterion,'' Nonlinearity {\bf 5}, 161 (1992).

\bibitem{devogel} R.~de~Vogelaere, ``On the structure
of symmetric periodic solutions of conservative
systems, with applications,'' in: {\it 
Contributions to the Theory of Nonlinear Oscillations},
Vol.~IV, ed. S.~Lefschetz, Princeton University Press,
Princeton, New Jersey (1958), p.53.

\bibitem{howard} J.E.~Howard and S.M.~Hohs, ``Stochasticity and
  reconnection in Hamiltonian systems,'' Phy.~Rev.~A {\bf
  29}, 418 (1984); J.E.~Howard and J.~Humphreys, 
``Nonmonotonic twist maps,'' Physica~D {\bf 80}, 256 (1995).

\bibitem{stix} T.H.~Stix, ``Current penetration and
plasma disruption,'' Phys.~Rev.~Lett. {\bf 36},
10 (1976).

\bibitem{wurm} A.~Wurm, ``Renormalization Group Applications in Area-Preserving
Nontwist Maps and Relativistic Quantum Field Theory,'' Ph.D. thesis,
The University of Texas at Austin, Austin (May 2002).

\bibitem{note2} There is a proof for the existence of smooth
  bifurcation curves for small values of $b$.\cite{delshams}

\bibitem{note3} For comments
 about the numerical accuracy see  Sec.~\ref{sssec:erroran}.

\bibitem{note4} Note that these equations are derived {\it assuming}
that the shearless curve scales at criticality. The validity of this
assumption is demonstrated {\it a posteriori} by 
Fig.~\ref{fig:torusatcrit}.

\bibitem{abad} J.J.~Abad and H.~Koch, ``A renormalization
group for Hamiltonians: numerical results,'' Nonlinearity
{\bf 11}, 1185 (1998); J.J.~Abad and H.~Koch, ``Renormalization and
periodic orbits for Hamiltonian flows,'' Comm.~Math.~Phys.
{\bf 212}, 371 (2000).

\bibitem{chandre3} C.~Chandre and H.R.~Jauslin, ``Renormalization-group
analysis for the transition to chaos in Hamiltonian systems,'' Phys.~Rep.
{\bf 365}, 1 (2002).

\bibitem{escande} D.F.~Escande and F.~Doveil, ``Renormalization
method for computing the threshold of the large-scale stochastic
instability in two degrees of freedom Hamiltonian systems,''
J.~Stat.~Phys. {\bf 26}, 257 (1981).

\bibitem{greene4} J.M.~Greene, ``How a swing behaves,''
Physica~D {\bf 18}, 427 (1986).

\bibitem{koch} H.~Koch, ``A renormalization group for
Hamiltonians, with applications to KAM tori,'' 
Ergod.~Th.~Dyn.~Sys. {\bf 19}, 475 (1999); H.~Koch, ``A
renormalization group fixed point associated with the breakup of
golden invariant tori,'' mp\_arc 02-175, Apr. 9 (2002).

\bibitem{mackay1} R.S.~MacKay, {\it Renormalization in Area
Preserving Maps}, Ph.D. thesis, Princeton (1982); R.S.~MacKay, ``A
  renormalization approach to invariant circles in area-preserving
  maps,'' Physica~D {\bf 7}, 283 (1983).

\bibitem{shenker} S.J.~Shenker and L.P.~Kadanoff, ``Critical
Behavior of a KAM surface: I. Empirical results,'' 
J.~Stat.~Phys. {\bf 27}, 631 (1982).

\bibitem{greene6} J.M.~Greene, ``The status of KAM theory
from a physicist's point of view,'' in: {\it Chaos in
Australia}, eds. G.~Brown and A.~Opie, World Scientific,
Singapore (1993), p.8.

\end{thebibliography}
\end{document}